\documentclass[aps,superscriptaddress]{revtex4}
\usepackage{amssymb,amsmath,epsfig}
\usepackage[colorlinks=true, pdfstartview=FitV, linkcolor=blue, citecolor=red, urlcolor=magenta, breaklinks=true]{hyperref}
\usepackage{graphicx}
\usepackage{epstopdf}
\usepackage[capposition=top]{floatrow}
\usepackage{subfig}

\usepackage{cleveref}
\newcommand{\appref}[1]{Appendix~\ref{#1}}

\usepackage{orcidlink}

\begin{document}


\title{Tachyonic AdS/QCD, Determining the Strong Running Coupling and $\beta$-function in both UV and IR Regions of AdS Space}

\author{Adamu Issifu \orcidlink{0000-0002-2843-835X}}
\email{ai@academico.ufpb.br}
\affiliation{Departamento de F\'isica, Instituto Tecnol\'ogico de Aeron\'autica, DCTA, 12228-900, S\~ao Jos\'e dos Campos, SP, Brazil}
\affiliation{Laborat\'orio de Computa\c c\~ao Cient\'ifica Avan\c cada e Modelamento (Lab-CCAM), S\~ao Jos\'e dos Campos, SP, Brazil}
\affiliation{CFisUC, Department of Physics, University of Coimbra, 3004-516 Coimbra, Portugal}

 \author{Elijah A. Abbey~\orcidlink{0000-0002-2599-7551}}
 \email{elijah@df.ufscar.br}
 \affiliation{Departamento de F\'isica, Universidade Federal de S\~ao Carlos,\\ 
  Caixa Postal 676, 13565-905 São Carlos, SP, Brazil}

\author{Francisco A. Brito~\orcidlink{0000-0001-9465-6868}}
\email{fabrito@df.ufcg.edu.br}
\affiliation{Departamento de F\'{\i}sica, Universidade Federal de Campina Grande
Caixa Postal 10071, 58429-900 Campina Grande, Para\'{\i}ba, Brazil}
\affiliation{Departamento de F\'isica, Universidade Federal da Para\'iba, 
Caixa Postal 5008, 58051-970 Jo\~ao Pessoa, Para\'iba, Brazil}

\begin{abstract} 
In this paper, we investigate the Quantum Chromodynamics (QCD)-like running coupling, $\alpha_s^{AdS}(Q^{2})$, and its associated $\beta$-function within a tachyonic Anti-de Sitter (AdS)/QCD framework. The $\text{AdS}_5$ bulk geometry is deformed through the introduction of a color dielectric function $G(\phi(z))$, associated with a tachyon field $\phi(z)$. This function governs the behavior of $\alpha_s^{AdS}(Q^{2})$ across all momentum scales by modifying the AdS background at both small and large values of the holographic coordinate $z$.In the ultraviolet (UV) regime (small $z$), the deformation is driven by free tachyons and reproduces features consistent with perturbative QCD. In contrast, in the infrared (IR) regime (large $z$), tachyon condensation dominates, yielding behavior characteristic of nonperturbative QCD. This construction enables a unified description of the running coupling and its $\beta$-function over the full range of momentum transfer $Q^2$, where $Q^2$ denotes the space-like momentum scale.
\end{abstract}

\maketitle
\pretolerance10000

\section{Introduction}

AdS/CFT correspondence \cite{correspondence-a,correspondence-b} has attracted considerable attention in recent years. This duality scenario enables us to establish a relation between quantum gravity on \(d+1\)-dimensional Anti-de Sitter (AdS) space and \(d\)-dimensional conformal field theory (CFT). In principle, we can calculate physical observables in strongly coupled gauge theory in terms of weakly coupled classical gravity theory. One of the most intriguing examples \cite{correspondence} of the duality is the correspondence between \(\text{SU}(N)\) \(\mathcal{N}=4\) supersymmetric Yang–Mills (SYM) theory and gravity. The correspondence is the basis for initiating holographic principles by 't Hooft \cite{correspondence1}. The gravitational dual is type IIB supergravity or string theory \cite{correspondence2}. This correspondence has proven to provide a supergravity explanation for field theories that display confinement and chiral symmetry-breaking characteristics \cite{Klebanov, Maldacena1}. The gauge theory dual to the AdS space has shown that the potential between two static color point particles has a Coulombic contribution to the net potential behavior \cite{Rey, Maldacena2}, indicating a deconfinement phase transition. Furthermore, the investigation presented in \citep{Witten} suggests that, when considering a compact boundary, AdS corresponds to the confinement phase while the black hole solution in AdS corresponds to the deconfinement phase. This interesting outcome is associated with the Hawking–Page transition \cite{Hawking}, where two distinct solutions to the Einstein equations—AdS space and the AdS black hole—are considered.

{Also, QCD phenomenology derived from holographic models has been applied successfully in studying several strong-interaction characteristics, such as light-flavor mesons \cite{Bernardini} and glueball phenomenology \cite{Bernardini1} using a more robust approach referred to as configurational entropy \cite{Gleiser, Gleiser1, Ferreira, Karapetyan:2019ran}.

In the AdS/CFT correspondence \cite{correspondence-a,correspondence-b, correspondence}, the effective string tension \(T_{\text{string}}\) and the string coupling \(g_s\) are related through the 't Hooft coupling \(\lambda := g^2 N\) as \(g_s = \lambda/(4\pi N)\) and \(T_{\text{string}} = \sqrt{\lambda}/(2\pi)\) \cite{Hoker, N-Beisert}. Thus, \(g_s = g^2/4\pi\), where \(g\) is the gauge coupling of the SYM theory. In the limit \(N \rightarrow \infty\), \(g_s \rightarrow 0\), so perturbative string theory is applicable. On the other hand, in the limit \(\lambda \rightarrow 0\) (weak gauge coupling), perturbation theory in terms of Feynman diagrams applies to the gauge theory \cite{Hoker, N-Beisert}. In the 't Hooft limit \(N \rightarrow \infty\), \(\lambda = \text{fixed}\), the spacetime curvature \(\mathcal{R}\) is much smaller than the string scale \(1/g_s\), so gravity provides a good approximation to the gauge theory. Otherwise, both theories are highly complex, making their equivalence nontrivial. Also, small curvature \(\mathcal{R}\) means a large AdS radius \(R\), since \(\mathcal{R} \sim 1/R^2\), with \(R = (4\pi g_s N)^{1/4} l_s\), where \(l_s\) is the string length \cite{correspondence}. The gauge and the string couplings are related by \(g_s \sim g^2\), and in the 't Hooft limit, the string coupling becomes so small that stringy effects decouple. In this limit, non‑planar interactions are removed, so we consider free strings on the \(\text{AdS}_5 \times \text{S}^5\) background. Additionally, large \(N\) does not directly correspond to QCD; however, in many instances, results for large \(N\) are qualitatively similar to those for \(N_c = 3\), where \(N_c\) is the number of colors in QCD. The 't Hooft parameter \(\lambda\) divides large-\(N\) QCD into two regions. In the limit \(\lambda \ll 1\), we have a perturbative theory where Feynman diagrams are used to calculate amplitudes. In the limit \(\lambda \gg 1\), we have strong coupling with non‑planar diagrams \cite{Ballon}. Also, light‑front holography establishes a relationship between the boost‑invariant light‑front wavefunctions and bound‑state amplitudes in AdS space \cite{light-front, tHooft:1993dmi}. Holographic QCD provides a suitable description of hadrons with known spectroscopic dynamical characteristics \cite{dilaton-profile, holographic-qcd}.

In this work, we deform the AdS space \cite{Polchinski,Polchinski1,Boschi,Boschi1,Brodskya,Brodsky1} with a Higgs‑like dimensionless \textit{color dielectric function} \(G(\phi)\) associated with a tachyonic potential \citep{Issifu,Issifu1,Issifu2,Issifu3}. The color dielectric function deforms the AdS space in the UV region in the presence of free tachyons (active tachyons), while the tachyon‑condensed color dielectric function \(G(\eta)\) (where \(\eta\) is the condensed tachyon field) also deforms the AdS space in the IR region, similar to the positive‑sign dilaton profile used in determining strong couplings in AdS/QCD. The AdS action intended for this study is formulated from the Dirac–Born–Infeld (DBI) action modified at a tachyonic vacuum, similar to Sen’s \(\text{AdS}_5\) tachyonic action \cite{Ashoke}. Firstly, we investigate the characteristics of both the strong running coupling \(\alpha_s(Q^2)\) and the associated \(\beta\)-function \(\beta(Q^2)\) by distorting the AdS space with the tachyonic \(G(\phi)\). Secondly, we examine \(\alpha_s(Q^2)\) and \(\beta(Q^2)\) by distorting the AdS space with \(G(\eta)\). We expect \(\alpha_s(Q^2)\) and \(\beta(Q^2)\) to behave similarly to perturbative QCD (pQCD) for the UV deformation of the AdS space. On the other hand, we expect AdS/QCD or nonperturbative QCD‑like behavior for the IR deformation of the AdS space. The results obtained from these regions will be compared with effective couplings determined from different observables, such as lattice QCD results, QCD phenomenology, and the \(g_1\) scheme, where \(\alpha_{g1}(Q^2)\) is extracted from the well‑measured Bjorken sum rule \cite{Adolph, Deur1, Bjorken, Bjorken:1969mm, gp1, Deur:2014vea, E143:1998hbs, Ackerstaff, HERMES:1998cbu, HERMES:2006jyl, E155:2000qdr, E155:1999pwm, E142:1996thl, E154:1997xfa}. This approach provides a new perspective on AdS/QCD, in which pQCD coupling constants can be determined directly from a UV deformation of the AdS space rather than from extrapolation from an IR deformation. Also, we study the parameter that controls the transition from pQCD to nonperturbative QCD.

We present a unified holographic framework in which a Higgs‑like tachyon potential deforms the AdS geometry through a single color dielectric function \(G(\phi)\). {In this approach, the deformation due to free tachyon modes governs the UV regime, yielding a decreasing coupling, \(\alpha_s^{\text{AdS}}(Q^2)\), that connects the IR non‑perturbative regime to the intermediate energy region where the gluon condensate changes sign, and the coupling approaches the perturbative domain.} Unlike models that extrapolate from the IR to the UV (e.g., see a comprehensive review in \cite{AdS-book}), this method derives both the perturbative and nonperturbative behaviors directly from the corresponding geometric deformations, establishing a continuous and unified description across all momentum scales without interpolation.

Additionally, we show that \(\alpha_s(Q^2)\) and \(\beta(Q^2)\) are related to scalar glueballs with mass \(m_\phi\) in the IR regime or the tachyonic mass scale $\tilde{m}_\phi$ in the UV regime, and discuss their effect. We discuss any \textit{Landau singularity} that may be observed in the UV region and propose how to deal with it in the model framework. The strong running coupling is a subject of active research due to its limited understanding in the low-momentum transfer region. Good knowledge of \(\alpha_s(Q^{2})\) at \(Q \rightarrow \infty\) is necessary to match the growing precision of hadron scattering experiments and to enhance the understanding of high‑energy unification of strongly interacting and electroweak theories. On the other hand, a precise understanding of \(\alpha_s(Q^{2})\) at \(Q \rightarrow 0\) on the scale of the proton mass enables us to understand hadron structure, confinement, and hadronization \cite{Olive, coupling, Altarelli, AdS-book} (and references therein).

This paper is organized as follows. In Sec.~\ref{sec1}, we review well‑known holographic QCD in two subsections: in Sec.~\ref{H} we review light‑front holography, and in Sec.~\ref{H1} we review AdS/CFT and holographic QCD. We set the basis for the study in Sec.~\ref{SG}; under this section, we discuss confinement at the tachyonic vacuum in Sec.~\ref{v}, and the tachyonic AdS/QCD action in Sec.~\ref{sec2}. We proceed to present the deformation of the AdS space in the UV region in Sec.~\ref{DA}, under which we discuss the strong running coupling in Sec.~\ref{DA1} and the \(\beta\)-function in Sec.~\ref{DA2}. In Sec.~\ref{D}, we study the deformation of the AdS space in the IR region, leading to the study of the associated strong running coupling in Sec.~\ref{D1} and \(\beta\)-function in Sec.~\ref{D2}. We finalize by presenting our analysis and conclusion in Sec.~\ref{A}, with the analysis in Sec.~\ref{A1} and the conclusion in Sec.~\ref{sec3}.

\section{Review of Holograhic QCD}\label{sec1}
\subsection{Light-Front Holography}\label{H}
Light-front (LF) quantization \cite{Brodsky2, Brodsky2a} provides suitable grounds for investigating the structure of the hadrons concerning quark and gluon degrees of freedom. The LF wavefunctions (LFWFs) are the relativistic generalization of the Schr\"odinger wavefunctions. They are determined at fixed time $\tau\,=\,x^+\,=\,x^0\,+\,x^3$ referred to as the LF time \cite{Dirac} instead of the 'usual' fixed time $t$. LF holography \cite{range-of-confinement, holographic-qcd, Brodsky3, Brodsky4, holographic-qcd2} relates the equations of motion in AdS space to QCD Hamiltonian formalism in physical spacetime quantized on LF at $\tau$. 
The correspondence allows for a direct relationship between hadronic amplitude $\Phi(z)$ in AdS space and LFWF $\phi_L(\zeta)$ describing constituent quarks and gluons that form the hadron, {with $\zeta$ the invariant impact LF variable}. The connection is a consequence of AdS/CFT correspondence, a theory established to be scale independent \cite{correspondence}. The term correspondence is applied because we can map a weakly interacting gravity-related theory in $d+1$-dimensional AdS space onto any strongly interacting CFT (such as SYM theory) in $d$-dimensions, giving rise to the name gauge/gravity duality. This mapping generates a relation between the $5^{th}$-dimension variable ($z$) of AdS space and $\zeta$ in physical spacetime. 

The mapping between $z$ and $\zeta$ was originally obtained by relating the electromagnetic current matrix element in AdS space \cite{Polchinski1} to the corresponding current matrix element in LF theory \cite{holographic-qcd, Abidin}. Also, we can establish a relationship between their energy-momentum matrix elements \cite{Brodsky4}; this demonstrates the consistency of holographic mapping to physical observables in LF. The resulting equation for meson bound states $q\bar{q}$ at $\tau$ has the form of single-variable relativistic Lorentz invariant Schr\"odinger equation
\begin{equation}\label{lf}
    \left(-\dfrac{d^2}{d\zeta^2}-\dfrac{1-4L^2}{4\zeta^2}+U(\zeta)\right)\phi_L(\zeta)=\mathcal{M}^2\phi_L(\zeta),
\end{equation}
where $\mathcal{M}$ is the mass, $L$ relative orbital angular momentum of $q$ (quark) and $\bar{q}$ (antiquark), $U(\zeta)\,=\,\kappa^4\zeta^2+2\kappa^2(L+S+1)$ is the confining potential in the soft-wall model with mass-scale $\kappa$ and spin $S$. A full derivation of \cref{lf} can be found in \cite{holographic-qcd2} and references therein. Therefore, when $\zeta$ is correctly defined, we can reduce the theory to a semi-classical approximation. In this case, the properties of the strong interacting dynamics are centered on an effective potential $U(\zeta)$ \cite{AdS-book}. The advantage of LF holography is its direct geometric interpretation of confinement and the natural identification of the holographic coordinate $\zeta$ with the invariant impact variable of the LF wavefunctions, enabling a unified description of hadron spectra and the running coupling \cite{holographic-qcd, holographic-qcd2}. Compared to other holographic approaches, it provides a simpler and more transparent framework to describe the scale dependence of $\alpha_s(Q^2)$ across both UV and IR regimes \cite{ads-beta-f, AdS-book}.

The mapping between the AdS coordinate \(z\) and the physical light-front variable \(\zeta\) is established by matching the AdS wave equation with Eq.~(1). For a two-particle bound state, the invariant impact variable \(\zeta = \sqrt{x(1-x)}\, b_\perp\), where \(x\) is the longitudinal momentum fraction and \(b_\perp\) the transverse separation, is identified with \(z\) up to a scale factor, i.e., \(\zeta = z\). This relation is derived by comparing electromagnetic current matrix elements in AdS space with their light-front counterparts, yielding \(z = \zeta\) in units where the AdS radius \(R = 1\) \cite{holographic-qcd2, Brodsky3, dilaton-profile}.

\subsection{AdS/CFT and Holographic QCD}\label{H1}
QCD is neither conformal nor supersymmetric. However, it displays approximate conformal behavior in the far ultraviolet (UV) region, where quark--gluon degrees of freedom dominate, and in the deep infrared (IR) region, where the relevant degrees of freedom are hadronic. To capture color confinement, conformal symmetry in $\text{AdS}_5$ space must therefore be broken. The AdS/CFT framework relies on the isomorphism between the Poincaré group together with the conformal group SO(4,2) and the isometry group of $\text{AdS}_5$ space, with the AdS metric providing the geometric foundation for constructing QCD-like models.
\begin{equation}\label{1}
ds^2=\dfrac{R^2}{z^2}\left( \eta_{\mu\nu}dx^{\mu}dx^{\nu}-dz^2 \right), 
\end{equation}
where $R$ is the AdS radius. The metric is invariant under scale changes in the fifth-dimensional variable $z\,\rightarrow\,\tilde{\lambda} z$ and the spacetime variable $x^{\mu}\,\rightarrow\,\tilde{\lambda} x^{\mu}$. The $z$ coordinate acts like a scaling variable in Minkowski space. Different values of $z$ correspond to different measures of {$4$-momentum} for which the hadron can be studied \cite{ads-beta-f}.

Some well-known functions that are introduced to break the $\text{AdS}_5$ space are dilaton profiles {$e^{\pm\phi_d(z)}$, with $\phi_d(z)$ the dilaton field, \cite{dilaton-profile,dilaton-profile1,Brodsky:2011zj}, or a warp factor $e^{A(z)}$, where $A(z)$ is any suitable function; commonly, logarithmic functions are used, i.e., $A(z)\,\sim\,\log(z)$.} Whichever function is chosen to distort the $\text{AdS}_5$ space leads to \textit{confinement} in uniquely different forms \cite{H-quark-p-ads}.

A particularly successful realization of this holographic approach is obtained by modifying the AdS$_5$ background with a positive-sign dilaton field, $\Phi(z) = +\kappa^2 z^2$. This profile leads to linear Regge trajectories, a confining potential, and a hadronic spectrum in excellent agreement with experimental data \cite{Brodsky1,ads-beta-f,dilaton-profile}. These changes induce an analytical \textit{color confinement}, leading to a nonperturbative AdS/QCD running coupling, $\alpha^{AdS}_s(Q^2)$. It is important to note that the understanding of the running coupling for pQCD is limited to short distances or high $Q^2$ regimes. In a similar sense, AdS/QCD with a positive dilaton background is based on long-distance interactions and the small $Q^2$ regime \cite{ads-beta-f}.

Analytically, the $\text{AdS}_5$ action bears resemblance to the general relativity action $S_{GR}\,\propto\,\int d^4x \sqrt{|\det g_{\mu\nu}|}\tilde{R}/G_N$, where $\tilde{R}$ is the Ricci scalar and $G_N$ is Newton's constant. In AdS/QCD, we replace the Ricci scalar with the gauge field as $\sqrt{\tilde{R}}\rightarrow F^{\mu\nu}$, ${G_N}$ is replaced by the gauge coupling {$\sqrt{G_N}\rightarrow g_5$}, and the modulus of the metric determinant $|\det g_{\mu\nu}|$ is replaced by the modulus of the $\text{AdS}_5$ metric determinant, $\sqrt{|\det g_{\mu\nu}|}\rightarrow\sqrt{|g_{AdS}|}e^{\kappa^2z^2}$, which includes the dilaton profile $e^{\kappa^2z^2}$ to distort the $\text{AdS}_5$ space; here, $\phi_d(z)=\kappa^2z^2$ (see \cite{AdS-book} for a more detailed analysis).

Therefore, the $\text{AdS}_5$ action takes the form
\begin{equation}\label{2}
S_{AdS}=-\dfrac{1}{4}\int d^4x\,dz\, \sqrt{|g_{AdS}|}e^{\kappa^2z^2}\dfrac{1}{g_{5}^2}F^{\mu\nu}F_{\mu\nu}.
\end{equation}
From Eq.(\ref{1}) the determinant of the metric is $\sqrt{|g_{AdS}|}=(R/z)^5$ and from the above expression, after introducing the dilaton profile, it turns into $\sqrt{|g_{AdS}|}\cdot e^{\kappa^2z^2}=(R/z)^5e^{\kappa^2z^2}$. The presence of the factor $e^{\kappa^2z^2}$ breaks the conformal symmetry while the $\text{AdS}_5$ geometry remains unchanged. Therefore, the expression $(R/z)^5e^{\kappa^2z^2}$, makes the separation between $e^{\kappa^2z^2}$ and the metric determinant, $(R/z)^5$, explicit. We emphasize that the dilaton field $\phi_d(z)$ enters the action as a multiplicative factor, $e^{\kappa^2 z^2}$, not as a modification of the metric determinant~\cite{dilaton-profile1, dilaton-profile, ads-beta-f}. The prefactor
\begin{equation}\label{2a}  
g_5^{-2}(z)=g_5^{-2}e^{\kappa^2z^2},
\end{equation}
restores the conformal symmetry in Eq.(\ref{2}) and leads to \textit{color confining} properties, in this case, $g_5^2$ has length dimension. Now, comparing $g_5$ to the coupling in the {$\mathcal{N}\,=\,4$ SYM theory $g$}, and mapping $z\,\rightarrow\,\zeta$, we can calculate the AdS/QCD $\alpha_s^{AdS}(Q^2)$ and the associated $\beta(Q^2)$ \cite{AdS-book}.

The combination of these models discussed above follows a coherent logical progression: Light‑Front holography provides the fundamental identification between the AdS coordinate $z$ and the physical impact variable $\zeta$; the soft‑wall model with a positive‑sign dilaton minimally implements confinement and linear Regge trajectories; and the DBI action (discussed below) with a tachyonic potential offers a string‑theoretic foundation that unifies the UV (free tachyon) and IR (condensed tachyon) deformations via a single color dielectric function. Each step builds on the previous one, resulting in a unified framework grounded in both phenomenology and string theory.

\subsection{AdS/CFT and Holographic QCD}\label{H1}
QCD is neither conformal nor supersymmetric. However, it displays approximate conformal behavior in the far ultraviolet (UV) region, where quark--gluon degrees of freedom dominate, and in the deep infrared (IR) region, where the relevant degrees of freedom are hadronic. To capture color confinement, conformal symmetry in $\text{AdS}_5$ space must therefore be broken. The AdS/CFT framework relies on the isomorphism between the Poincaré group together with the conformal group SO(4,2) and the isometry group of $\text{AdS}_5$ space, with the AdS metric providing the geometric foundation for constructing QCD-like models.
\begin{equation}\label{1}
ds^2=\dfrac{R^2}{z^2}\left( \eta_{\mu\nu}dx^{\mu}dx^{\nu}-dz^2 \right), 
\end{equation}
where $R$ is the AdS radius. The metric is invariant under scale changes in the fifth-dimensional variable $z\,\rightarrow\,\tilde{\lambda} z$ and the spacetime variable $x^{\mu}\,\rightarrow\,\tilde{\lambda} x^{\mu}$. The $z$ coordinate acts like a scaling variable in Minkowski space. Different values of $z$ correspond to different measures of {$4$-momentum} for which the hadron can be studied \cite{ads-beta-f}.

Some well-known functions that are introduced to break the $\text{AdS}_5$ space are dilaton profiles {$e^{\pm\phi_d(z)}$, with $\phi_d(z)$ the dilaton field, \cite{dilaton-profile,dilaton-profile1,Brodsky:2011zj}, or a warp factor $e^{A(z)}$, where $A(z)$ is any suitable function; commonly, logarithmic functions are used, i.e., $A(z)\,\sim\,\log(z)$.} Whichever function is chosen to distort the $\text{AdS}_5$ space leads to \textit{confinement} in uniquely different forms \cite{H-quark-p-ads}.

A particularly successful realization of this holographic approach is obtained by modifying the AdS$_5$ background with a positive-sign dilaton field, $\Phi(z) = +\kappa^2 z^2$. This profile leads to linear Regge trajectories, a confining potential, and a hadronic spectrum in excellent agreement with experimental data \cite{Brodsky1,ads-beta-f,dilaton-profile}. These changes induce an analytical \textit{color confinement}, leading to a nonperturbative AdS/QCD running coupling, $\alpha^{AdS}_s(Q^2)$. It is important to note that the understanding of the running coupling for pQCD is limited to short distances or high $Q^2$ regimes. In a similar sense, AdS/QCD with a positive dilaton background is based on long-distance interactions and the small $Q^2$ regime \cite{ads-beta-f}.

Analytically, the $\text{AdS}_5$ action bears resemblance to the general relativity action $S_{GR}\,\propto\,\int d^4x \sqrt{|\det g_{\mu\nu}|}\tilde{R}/G_N$, where $\tilde{R}$ is the Ricci scalar and $G_N$ is Newton's constant. In AdS/QCD, we replace the Ricci scalar with the gauge field as $\sqrt{\tilde{R}}\rightarrow F^{\mu\nu}$, ${G_N}$ is replaced by the gauge coupling {$\sqrt{G_N}\rightarrow g_5$}, and the modulus of the metric determinant $|\det g_{\mu\nu}|$ is replaced by the modulus of the $\text{AdS}_5$ metric determinant, $\sqrt{|\det g_{\mu\nu}|}\rightarrow\sqrt{|g_{AdS}|}e^{\kappa^2z^2}$, which includes the dilaton profile $e^{\kappa^2z^2}$ to distort the $\text{AdS}_5$ space; here, $\phi_d(z)=\kappa^2z^2$ (see \cite{AdS-book} for a more detailed analysis).

Therefore, the $\text{AdS}_5$ action takes the form
\begin{equation}\label{2}
S_{AdS}=-\dfrac{1}{4}\int d^4x\,dz\, \sqrt{|g_{AdS}|}e^{\kappa^2z^2}\dfrac{1}{g_{5}^2}F^{\mu\nu}F_{\mu\nu}.
\end{equation}
From Eq.(\ref{1}) the determinant of the metric is $\sqrt{|g_{AdS}|}=(R/z)^5$ and from the above expression, after introducing the dilaton profile, it turns into $\sqrt{|g_{AdS}|}\cdot e^{\kappa^2z^2}=(R/z)^5e^{\kappa^2z^2}$. The presence of the factor $e^{\kappa^2z^2}$ breaks the conformal symmetry while the $\text{AdS}_5$ geometry remains unchanged. Therefore, the expression $(R/z)^5e^{\kappa^2z^2}$, makes the separation between $e^{\kappa^2z^2}$ and the metric determinant, $(R/z)^5$, explicit. We emphasize that the dilaton field $\phi_d(z)$ enters the action as a multiplicative factor, $e^{\kappa^2 z^2}$, not as a modification of the metric determinant~\cite{dilaton-profile1, dilaton-profile, ads-beta-f}. The prefactor
\begin{equation}\label{2a}  
g_5^{-2}(z)=g_5^{-2}e^{\kappa^2z^2},
\end{equation}
restores the conformal symmetry in Eq.(\ref{2}) and leads to \textit{color confining} properties, in this case, $g_5^2$ has length dimension. Now, comparing $g_5$ to the coupling in the {$\mathcal{N}\,=\,4$ SYM theory $g$}, and mapping $z\,\rightarrow\,\zeta$, we can calculate the AdS/QCD $\alpha_s^{AdS}(Q^2)$ and the associated $\beta(Q^2)$ \cite{AdS-book}.

The combination of these models discussed above follows a coherent logical progression: Light‑Front holography provides the fundamental identification between the AdS coordinate $z$ and the physical impact variable $\zeta$; the soft‑wall model with a positive‑sign dilaton minimally implements confinement and linear Regge trajectories; and the DBI action (discussed below) with a tachyonic potential offers a string‑theoretic foundation that unifies the UV (free tachyon) and IR (condensed tachyon) deformations via a single color dielectric function. Each step builds on the previous one, resulting in a unified framework grounded in both phenomenology and string theory.

\section{The Model}\label{SG}
\subsection{Confinement at Tachyonic Vacuum}\label{v}
We start from the Dirac--Born--Infeld (DBI) action~\cite{Born,Gibbons,Dirac1,Polchinskia,Polchinski:1996na,Taylor} and extend it by including a tachyonic field near the tachyonic vacuum. In this regime, bosonic open strings effectively behave as closed strings without D$p$-branes, as shown in string field theory: their endpoints are connected by flux lines on the D$p$-brane worldvolume, forming closed strings~\cite{Sent1,Sen1,Sen:1998ki,Sen:1998sm,Recknagel,Polchinski:1994my,Callan:1994ub,Sen2,Harvey,Majumder:2000tt,Fendley:1994rh,Kostelecky,Sen,Zwiebach2,Kostelecky1,Moeller:2000jy,Harvey:2000tv,Moeller:2000xv,Taylor:2000ek,Berkovits:2000zj,Berkovits,Harvey1,Gopakumar,Minahan:2000tf,Gerasimov:2000zp,Harvey:2000jb,Dasgupta:2000ft,Kutasov:2000qp,Ghoshal:2000gt}.

From a quantum field theory viewpoint, D$p$-branes do not vanish completely at the tachyonic vacuum but instead dissolve into lower-dimensional branes, which connect open-string endpoints and act as sources of color charge. In this limit, the negative tachyon potential $V(\phi)$ exactly cancels the D$p$-brane tension $\varepsilon_p$, i.e.
\begin{equation}
    V(\phi) + \varepsilon_p = 0,
\end{equation}
leaving no net energy cost for rearranging strings on the residual lower-dimensional branes and eliminating open string excitations~\cite{Sen, Zwiebach2, Kostelecky, Berkovits, Witten2, Seiberg:1999vs, Sent2, Bigazzi}.  

In the present study, the relevant degrees of freedom are \textit{valence gluons}. Consequently, the closed strings formed at the tachyonic vacuum are treated within the \textit{color flux-tube picture} of QCD, consistent with \textit{chromoelectric flux confinement} (see Ref.~\cite{Issifu:2022pif} and references therein).

The DBI action takes the form
\begin{equation}\label{t}
    S_{\text{DBI}}=-T_p\int d^{p+1}\xi e^{-\phi_d}\sqrt{-\det(\eta_{\mu\nu}+(2\pi\alpha')F_{\mu\nu})},
\end{equation}
where $F_{\mu\nu}\,=\,\partial_\mu A_\nu-\partial_\nu A_\mu$ is the abelian gauge field strength living on the worldvolume of the D$p$-brane, D$p$-brane tension
\begin{equation}\label{tt2}
\tau_p=\dfrac{T_p}{g_s}=\dfrac{1}{g_s\sqrt{\alpha'}(2\pi\sqrt{\alpha'})^p},
\end{equation}
string tension 
\begin{equation}
T_{\text{string}}=\dfrac{1}{2\pi\alpha'},
\end{equation}
and string coupling $g_s=e^{\langle\phi_d\rangle}$. To understand the strong interaction dynamics of the open strings on the D$p$-brane worldvolume, we derive the dimensionally reduced $\text{U}(1)$ Yang-Mills theory \cite{Taylor1, Banks} by small perturbation in Eq.(\ref{t}), which yields
\begin{equation}\label{ymt1}
S_{YM}=-\tau_pV_p-\dfrac{1}{4g_{YM}^2}\int d^{p+1}\xi F_{\mu\nu}F^{\mu\nu} +{\cal O}(F^4).
\end{equation}  
Here, $V_p$ is the $p$-brane worldvolume, which can be ignored subsequently, and $g_{YM}$ is the Yang-Mills coupling given by,
\begin{equation}\label{ymt2}
g_{YM}^2=\dfrac{1}{(2\pi\alpha')^2\tau_p}=\dfrac{g_s}{\sqrt{\alpha'}}\left(2\pi\sqrt{\alpha'}\right)^{p-2},
\end{equation}
hence, the relation $g_{YM}^2\,\sim\,g_s$, that forms the basis for the gauge/gravity duality is established. 
Since the mass of the string is proportional to its length, {the closer the branes, the less massive the attached strings} may be \cite{Witten1, Taylor}. Therefore, for stacks of $N$ D$p$-branes, we have $\text{U}(1)^N$ gauge groups on its worldvolume which can be decomposed as $\text{U}(N)=\text{SU}(N)\times\text{U}(1)$. That can further be decomposed into standard model symmetry depending on the number of D$p$-branes crossing each other \cite{Lust, Chatzistavrakidis:2011gs, Ibanez:2001nd, Antoniadis:2002qm}.
Now we modify Eq.(\ref{t}) with a tachyon potential $V(\phi)$ \cite{Garousi} to induce color confinement, with $\phi$ the tachyon field
\begin{align}\label{tb}
    S&=-T_p\int d^{p+1}\xi e^{-\phi_d}V(\phi)\sqrt{-\det(\eta_{\mu\nu}+(2\pi\alpha')F_{\mu\nu})}\nonumber\\
    &=-\tau_p\int d^{p+1}\xi V(\phi)\left[1+\dfrac{1}{4}(2\pi\alpha')^2F_{\mu\nu}F^{\mu\nu}+{\cal O}(F^4)\right],
\end{align}
in the last step, we considered slowly varying tachyons with a constant dilaton field $\langle\phi_d\rangle$. At the minimum of the potential $V(\phi_0)$ the action disappears \cite{Sent2,Sent1,Recknagel}. Also, the potential is symmetric under $\phi_0\,\rightarrow\,-\phi$, with maximum at $\phi\,=\,0$ and minimum at $\phi_0\,=\,\pm a$.  We will disregard the first term and consider only the gauge fields propagating on the D$p$ worldvolume for the analysis, because the gauge field is enough for the current study. The background dilaton field $\phi_d(z)$ is incorporated into the effective color dielectric function via $G(\phi,z) = e^{-\phi_d(z)} (2\pi\alpha')^2 V(\phi)$. Evaluated on-shell, this defines a scale-dependent coupling $G(\phi(z))$. Thus, the dilaton controls the $z$-dependence of the gauge coupling without modifying the AdS metric. This is consistent with standard soft-wall AdS/QCD treatments, where conformal symmetry breaking is encoded in background fields rather than in the geometry. 
Adopting the conventional normalization in string units, $(2\pi\alpha')^2 = 1$ and $\langle\phi_d\rangle$ the relation simplifies to $G(\phi) = V(\phi)$, rendering $G(\phi)$ dimensionless. Consequently, we have,
\begin{equation}\label{td}
   \mathcal{L}=-\dfrac{1}{4g_s}G(\phi)F_{\mu\nu}F^{\mu\nu},
\end{equation}
for $T_p\,=\,1$. Introducing the kinetic energy component and the tachyon potential into the last step of Eq.(\ref{tb}), we recover the Lagrangian density in references \cite{Issifu, Issifu1, Issifu2, Issifu3, dielectric-f} used in studying color confinement.  

Although the DBI gauge field is Abelian, the non-Abelian dynamics of QCD are captured by the holographic background \cite{Gursoy, Gursoy:2007er}. In AdS/QCD, the running coupling is obtained from the two-point function of the gauge field strength, which reduces to a scalar problem in the isotropic background of the metric, dilaton, and tachyon \cite{Ashoke}. The tachyon potential and color dielectric function encode the non-perturbative gluon dynamics, including confinement and the running of $\alpha_s(Q^2)$, in the same way as standard AdS/QCD models where an Abelian vector field \cite{ads-beta-f, dilaton-profile} suffices to describe hadronic form factors and the QCD coupling. Thus, the Abelian approximation \cite{dilaton-profile1, Grigoryan:2007vg} is fully adequate for computing $\alpha_s(Q^2)$, as the non-Abelian structure is already incorporated through the background fields.
 }

\subsection{Tachyonic AdS/QCD Action} \label{sec2}
Considering $5$D holographic QCD models, the background flavor branes are  D$4$-$\bar{\text{D}}4$ systems \cite{Bigazzi1, Gursoy, Gursoy:2008za, Gursoy:2007er}. In this study, we consider open strings with their endpoints on a D$p$-brane worldvolume, so we consider a background D$4$-brane. Thus, we can express the $\text{AdS}_5$ action similar to the one proposed in \cite{Ashoke} adopted in \cite{Iatrakis} for studying hadronic properties in the light of AdS/QCD, from Eq.(\ref{tb}) we can express
\begin{equation}\label{11}
    S=-\dfrac{1}{4g_5^2}\int d^4xdzG(\phi(z))F^{\mu\nu}F_{\mu\nu},
\end{equation}
where $G(\phi(z))$ is only a function of $z$ and $g_s\,\rightarrow\,g_5^2$ has the dimension of length. Additionally, tachyonic corrections were introduced in Ref.~\cite{Barbosa} to show equivalence in AdS/QCD and tachyonic AdS/QCD \cite{Bajnok} in $4$D holographic dual to $5$D weakly coupled gravity using a similar action.  The $G(\phi)$ is carefully defined to induce confinement and asymptotic freedom in separate energy regimes. Again, we demonstrate that the \textit{tachyonic color dielectric function}, $G(\phi)$, deforms the $\text{AdS}_5$ metric at the asymptotic $\text{AdS}_5$ bulk in the UV limit ($z\,\rightarrow\,0$) corresponding to deconfinement transition associated with $\text{AdS}_5$ black holes \cite{Hawking}. On the other hand, deforming the $\text{AdS}_5$ metric with \textit{condensed tachyonic color dielectric function} $G(\eta)$ leads to IR deformation of the $\text{AdS}_5$ bulk leading to the phenomenon of \textit{color confinement} in the IR regime ($z\,\rightarrow\,\infty$), similar to deforming the 'usual' $\text{AdS}_5$ space with positive dilaton background \cite{Brodsky1}. Here, the confining transition corresponds to color singlet glueball states with {$N^{0}$} degree of freedom, whilst the deconfinement transition corresponds to the state where the gluons are free with {$N^{2}$} degrees of freedom \cite{Witten, Ballon}.

Moreover, in the model framework, the tachyon field $\phi$ is dual to the scalar glueball operator $O = \mathrm{Tr}\,G_{\mu\nu} G^{\mu\nu}$, as presented in Eqs.~(\ref{13c}) to (\ref{opr}). At the false vacuum ($\phi = 0$), $\langle O \rangle \propto V(0) = \Lambda a^4/4$, while at the true vacuum ($\phi = \pm a$), the tachyon condenses, yielding $\langle O \rangle \propto a^4$, which encodes the gluon condensate in the confined phase. Fluctuations $\eta$ around $\phi_0 = \pm a$ correspond to physical scalar glueball states with quantum numbers $0^{++}$.

\section{Deformation of the AdS Space in the UV Region}\label{DA}
\subsection{Strong Running Coupling}\label{DA1}
The strong running coupling $\alpha_s^{AdS}(Q^2)$ can be determined by defining a new coupling  $g_5(z)$ with dependence on $z$, that restores the conformal symmetry in Eq.(\ref{11}) (this has been discussed extensively in Sec.~\ref{H1}),
\begin{equation}\label{12}
g_5^{-2}(z)=g_5^{-2}G(\phi(z)),
\end{equation}
this approach is similar to the one used in determining the QCD running coupling of AdS/QCD \cite{AdS-book}. The deformation of the conformal symmetry in Eq.(\ref{11}) is relevant for obtaining running coupling and a nonzero $\beta$-function because conformal symmetry leads to constant coupling and zero $\beta$-function, similar to what is observed in classical QCD without quantum correction. The scale-dependent gauge field depicted by the expression above is equivalent to gauge field strength renormalization in the QCD theory. Here, the gauge field is scaled as $F^{\mu\nu}\,\rightarrow\, G^{1/2}F^{\mu\nu}$ to induce QCD-like properties. 

In 'regular' QCD theory; we can rescale the gluon field as $A^\mu\,\rightarrow\,\tilde{\lambda} A^\mu$ which leads to $G^{\mu\nu}\,\rightarrow\,\tilde{\lambda} G^{\mu\nu}$, {with $G^{\mu\nu}$ the non-abelian field strength}, in the QCD Lagrangian density, $\mathcal{L}_{\text{QCD}}$, this is accompanied by rescaling the coupling strength $g\,\rightarrow\,\tilde{\lambda}^{-1}g$. Thus, the renormalization of the physical QCD coupling  $g_{phys}\,=\,Z^{1/2}_3g_0$ \cite{Nayak}, { with $Z_3$ the gluon propagator renomalization factor and $g_0$ the bare coupling} in UV-regulated theory similar to renormalization of gauge field and gauge field strength: $A_{ren}^\mu\,=\,Z^{-1/2}_3A_0^\mu$, $G^{\mu\nu}_{ren}\,=\,Z_3^{-1/2}G_0^{\mu\nu}$ resulting into a rescaled Lagrangian $\mathcal{L}_{\text{QCD}}^{ren}\,=\,Z_3^{-1}\mathcal{L}_{\text{QCD}}^{0}\,=\,(g_{phys}/g_0)^{-2}\mathcal{L}_{\text{QCD}}^{0}$. All quantities with subscript or superscript '$0$' are bare and the ones with '\textit{ren}' are their renormalized counterparts. A similar idea is applied in AdS/QCD calculations, where the regularized coupling is defined in terms of a suitable function that induces the QCD-like properties.

The nonconformal dynamics, confinement, and deconfinement transitions associated with QCD will now be investigated through $g_5(z)$ where $g_5(z)\,\rightarrow\,g_{YM}(\zeta)$ --- see Eq.(\ref{ymt1}) for the comparison with YM theory. Therefore, we define $\text{AdS}_5$ coupling constant, $\alpha^{AdS}_s\,\equiv\,g_5^2/4\pi$, which requires $Q^2$-dependence to fully describe the dynamics of the particles in both UV and the IR domains.  
What makes the $G(\phi)$ interesting is that it distorts the $\text{AdS}_5$ space curvature in two different forms. One in the UV regime and the other in the IR regime, so perturbation and nonperturbative characteristics of the QCD-like coupling can be investigated in a single framework \cite{ads-beta-f}. The $\alpha_s^{AdS}$ can be written in terms of the $G(\phi)$ as
\begin{equation}\label{12a}
\alpha_s^{AdS}(\zeta)=\dfrac{g^2_{YM}(\zeta)}{4\pi}\propto G(\phi(\zeta))^{-1}.
\end{equation}   
It is necessary to identify the nature of $G(\phi(\zeta))$ {in order to compute} $\alpha^{AdS}_s(Q^2)$ and $\beta(Q^2)$. 
The form of $G(\phi(\zeta))$ has been proposed in \cite{Issifu:2022pif} and used in investigating the phenomenon of (de)confinement of glueballs. As we have demonstrated in Sec.~\ref{v}, the dimensionless $G(\phi(\zeta))$ is directly associated with the tachyon potential $V(\phi)$ for strings with $T_{\text{string}}\,=\,1$ corresponding to $(2\pi\alpha')^2\,=\,1$ we can express
\begin{equation}\label{13}
V(\phi(\zeta))=G(\phi(\zeta))=\dfrac{\Lambda}{4}[\phi(\zeta)^2-a^2]^2.
\end{equation}
We will use this function to induce confinement and deconfinement transitions. In the string-theory normalization $(2\pi\alpha')^2 = 1$ \cite{Ashoke} (see below \cref{tb}), the tachyon field $\phi$ is dimensionless~\cite{Garousi,Tong:2009np}. Consequently, in the potential given in \cref{13}, the parameters $\Lambda$ and $a$ are dimensionless constants. With the linear ansatz $\phi(z) = \alpha z$, the parameter $\alpha$ has dimensions of energy (since $z$ has dimensions of energy$^{-1}$), and the tachyon decay constant is $f_\alpha = 1/\alpha$ (with dimensions of energy$^{-1}$). The function $G(\phi(\zeta))$ is also dimensionless, as required. 
The linear ansatz $\phi(z) = \alpha z$ \cite{Boschi, Boschi1, Polchinski, Polchinski1} is the simplest deformation that preserves the AdS geometry at $z \to 0$ and introduces a single scale $\alpha$ (the tachyon decay constant). Motivated by the soft‑wall model \cite{Brodsky1, Brodskya, AdS-book}, it captures the UV behavior to leading order and describes the transition region up to a few GeV, consistent with the $g_1$ effective charge \cite{Deur1, Bjorken, Bjorken:1969mm, Deur:2014vea, gp1}. More refined profiles reproducing full pQCD running could be explored in future work, beyond this leading-order study. Therefore, 
\begin{equation}\label{13a}
    G(\phi)=\dfrac{\Lambda}{4}[(\zeta\alpha)^2-a^2]^2.
\end{equation}
The scale anomaly in QCD induces a nonvanishing energy-momentum trace tensor, given as
\begin{equation}\label{13c}
    \left\langle\Theta^\mu_\mu\right\rangle=-\dfrac{9}{8}\left<\dfrac{\alpha_s}{\pi}G^{\mu\nu}G_{\mu\nu}\right>,
\end{equation}
where $\alpha_s$ is the strong coupling and $G^{\mu\nu}$ is the non-abelian gauge field strength. Using parton-hadron duality principle \cite{Schechter,Carter,Campbell,Kochelev1} we can establish a relation between $\left\langle\Theta^\mu_\mu\right\rangle$ and $\Lambda$ as
{\begin{align}\label{13b}
    \left\langle \Theta^\mu_\mu\right\rangle&=\left(4V(\phi)-\phi\dfrac{dV}{d\phi}\right)\Big\vert_{\phi_0=0}\nonumber\\
     &=4V(\phi_0)=\Lambda a^4.
\end{align}}
Equating Eqs.(\ref{13c}) and (\ref{13b}) yields,
\begin{equation}\label{opr}
    \Lambda=\dfrac{9\left<\alpha_sG^{\mu\nu}G_{\mu\nu}\right>}{8\pi a^4},
\end{equation}
hence, the coupling $\Lambda$, is proportional to the gluon condensate \cite{Kharzeev}.
Now, calculating the physical running coupling measured on $Q^2$-scale, we use the Mellin transform. The Mellin transform relating the holographic coordinate $\zeta$ to momentum space follows from the AdS/CFT correspondence, where $\zeta$ is dual to the renormalization scale $\mu \sim 1/\zeta$ \cite{Mack:2009mi}. The bulk-to-boundary propagator for a scalar field of mass $\tilde{m}_\phi$ in AdS$_5$ introduces a weight $\zeta^{Q^2/|\tilde{m}_\phi^2| - 1}$. 

The integrand in \cref{13a} contains a double pole on the real axis at $\zeta_0 = a/\alpha$, rendering the naive integral ill-defined. The physical coupling $\alpha_s^{\mathrm{AdS}}(Q^2)$ is therefore defined by analytic continuation, implemented through the prescription $\zeta \to \zeta + i\varepsilon$ with $\varepsilon \to 0^+$, which shifts the pole infinitesimally off the contour. The resulting expression is evaluated in the sense of distributions, reducing to a Cauchy principal value integral after a single integration by parts, as detailed in \appref{app:mellin}.
Carrying out the Mellin transform yields
\begin{align}\label{14}
\alpha^{\mathrm{AdS}}_s(Q^2)
&\sim \int^{\infty}_{0}\alpha^{\mathrm{AdS}}_s(\zeta)\,\zeta^{Q^2/|\tilde{m}_\phi^2|-1}\, d\zeta \nonumber\\
&=\dfrac{\sigma\pi}{\Lambda a^4 |\tilde{m}_\phi^2|}\left( -\dfrac{\alpha^2}{a^2}\right)^{-Q^2/2|\tilde{m}_\phi^2|}(2|\tilde{m}_\phi^2|-Q^2)\csc\!\left( \dfrac{\pi Q^2}{2|\tilde{m}_\phi^2|}\right)\nonumber\\
&=\dfrac{\sigma\Lambda\pi}{|\tilde{m}_\phi^6|}\left(\dfrac{|\tilde{m}_\phi^2|}{\Lambda}f_\alpha^2\right)^{Q^2/2|\tilde{m}_\phi^2|}(2|\tilde{m}_\phi^2|-Q^2)\csc\!\left(\dfrac{\pi Q^2}{2|\tilde{m}_\phi^2|}\right).
\end{align}
The trigonometric structure arises from the Euler reflection formula $\Gamma(p)\Gamma(1-p)=\pi/\sin(\pi p)$ underlying the Mellin transform of functions with poles on the integration contour. In particular, the cosecant factor emerges from the principal value integral after combining contributions at shifted arguments, as shown in \appref{app:mellin}. To avoid ambiguity between UV and IR mass scales, we distinguish the tachyonic instability scale at the false vacuum from the physical glueball mass. At $\phi=0$, the potential curvature defines $\tilde{m}_\phi^2 = V''(0) = -\Lambda a^2 < 0$, which characterizes the UV deformation. Since a tachyon has spacelike momentum, $p^2 = \tilde{m}_\phi^2 < 0$, physical observables depend only on its magnitude. We therefore define the positive scale $m_{\phi,\mathrm{UV}}^2 \equiv |\tilde{m}_\phi^2| = \Lambda a^2$, which enters the running coupling. This is required to keep the Mellin representation (\cref{14}) real, as using $\tilde{m}_\phi^2 < 0$ would lead to an unphysical complex coupling. In contrast, the physical glueball mass arises from fluctuations about the true vacuum $\phi = \pm a$, where $m_\phi^2 = V''(\pm a) = 2\Lambda a^2 > 0$ (which will be discussed in detail in \cref{D}). The two scales are related by $m_\phi = \sqrt{2}\,|\tilde{m}_{\phi}|$. Numerically, we take $|\tilde{m}_{\phi}| = 0.86~\mathrm{GeV}$ and $m_\phi = 1.73~\mathrm{GeV}$ as independent lattice-informed inputs, with their ratio consistent with the model expectation at leading order.

We substituted (\cref{12a}) and divided $Q^2$ by $|\tilde{m}^2_\phi|$, in the integral for dimensional consistency. We will demonstrate below that the mass scale is related to the scale $Q_\Lambda$ at which pQCD breaks down in the model framework. The constant $\sigma$ originates from the string tension, $T_{\rm string}=1/(2\pi\alpha')$ and $\sigma=T_{\rm string}^2$, where $T_{\rm string}$ is the D$p$-brane string tention. In string units where \((2\pi \alpha')^2 = 1\), all dimensionful quantities are measured in terms of the string energy scale \(E_s = 1/\sqrt{\alpha'} = \sqrt{2\pi}\) (in units of the string length), so that \(\sigma = T_{\rm string}^2 = 1/(2\pi \alpha')^2 = 1\) is consistent. Upon restoring physical units, one identifies \(E_s\) with a physical mass scale; since the only explicit scale in \cref{14} is \(|\tilde{m}_\phi|\), we set \(E_s \equiv |\tilde{m}_\phi|\), which yields \(\sigma = E_s^4 = |\tilde{m}_\phi^{\,4}|\) in \(\text{GeV}^4\).

We showed in (\cref{tb}) and stated below it that $G(\phi)$ is a dimensionless function associated with the tachyon potential through the Regge slope, so the proportionality constant serves as a dimensional correction to the expression since dimensionful quantities such as the glueball mass and the energy scale $Q^2$ were introduced. It should be noted that in the tachyon potential, $a$ and $\phi$ have dimensions of energy, but $G(\phi)$ is dimensionless. For mathematical convenience, we set the dimensionless coupling constant, $\Lambda$, to unity, $\Lambda\,=\,1$, and also recall $\sigma=|\tilde{m}_\phi^4|$ for dimensional consistency. Therefore, Eq.(\ref{14}) simplifies to 
\begin{equation}\label{15}
\alpha^{AdS}_s(Q^2)=\dfrac{\sigma\pi}{|\tilde{m}_\phi^6|}\left({|\tilde{m}_\phi^2|f_\alpha^2}\right)^{Q^2/2|\tilde{m}_\phi^2|} (2|\tilde{m}_\phi^2|-Q^2)\csc\left(\dfrac{\pi Q^2}{2|\tilde{m}_\phi^2|}\right).
\end{equation}
We emphasize that \cref{15} applies in the regime $Q^2 \lesssim |\tilde{m}_\phi^2|$. For $Q^2 \to \infty$, it falls as $1/Q^2$, characteristic of a tree-level bulk result. The logarithmic running of perturbative QCD would require quantum corrections in the bulk or a running tachyon profile $\phi(z)$. Thus, the model describes the IR--intermediate regime, while the deep UV requires additional physics, such as the dynamically generated gluon mass introduced in \cref{A1}.

\begin{figure}[H]
\centering
  \subfloat[Left Panel]{\includegraphics[scale=0.5]{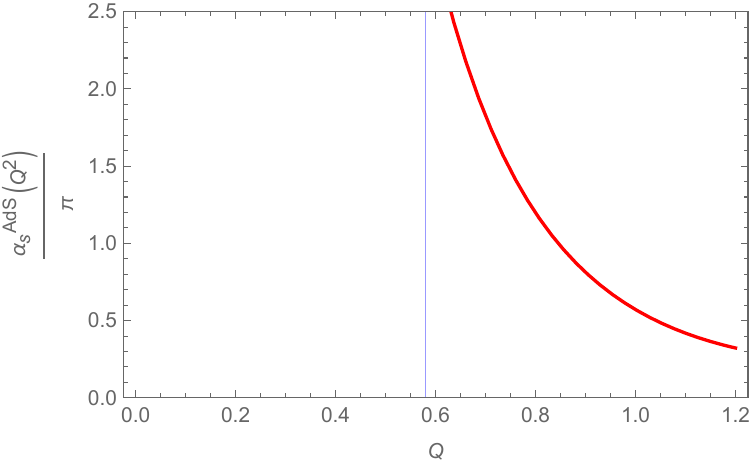}}
  \qquad
  \subfloat[Right Panel]{\includegraphics[scale=0.5]{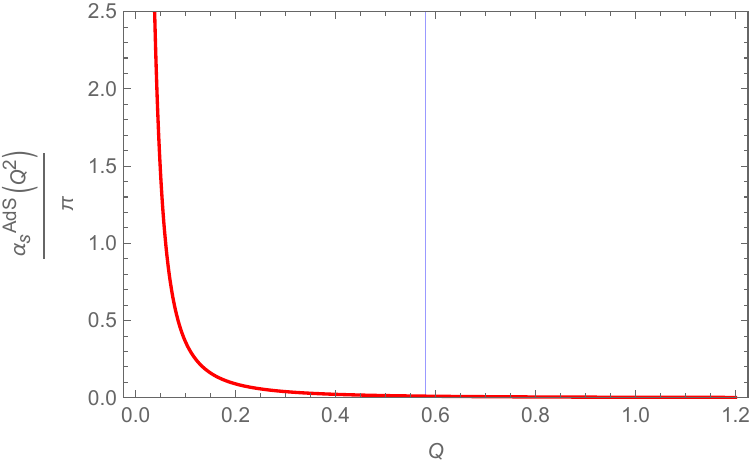}}
 \caption{A graph of $\alpha^{AdS}_s(Q^2)/\pi$ against $Q$ for $|\tilde{m}_\phi|\,=\,0.86\,\text{GeV}$ (left) and $|\tilde{m}_\phi|\rightarrow\infty$ (right). We compare the graph (left) with the ones contained in Refs.~\cite{Brodskya, AdS-book} for $\alpha_{g1}(Q)/\pi$ (pQCD) and other experimental data in \cite{Deur}, and we observe the expected decrease of $\alpha_s$ at large $Q$. We present results for both the lightest glueball mass (left) and the heaviest possible glueball mass (right). However, the theory governing the study is consistent with light glueball masses, so the graph in the right panel is for analytical purposes and may not have any physical implications. We observe that $\alpha_s^{AdS}(Q^2)$ decreases sharply with increasing $Q$; on the other hand, an increase in $|\tilde{m}_\phi|\,\rightarrow\,\infty$ moves the graph more towards a small $Q$ region beyond the Landau pole, $Q_\Lambda$, (vertical blue line)  and also falls quickly at large $Q$. Thus, heavy glueballs are more likely to be tightly bound than lighter ones.}
   \label{Fig1}
\end{figure}

{We plot the leading order of the perturbative result of the QCD running coupling given by 
\begin{equation}
   \alpha_s^{pQCD}(Q^2)=\dfrac{4\pi}{\beta_0\ln\left(\dfrac{Q^2}{\Lambda^2_{QCD}}\right)}
\end{equation}
where $\Lambda_{QCD}$ is the QCD scale,
\begin{equation}
    \beta_0=11-\dfrac{2}{3}n_f,
\end{equation}
is the first term of the $\beta$-series, with $n_f$ the number of quark flavors active at scale $Q^2$.
}
\begin{figure}[H]
  \centering
 \includegraphics[scale=.6]{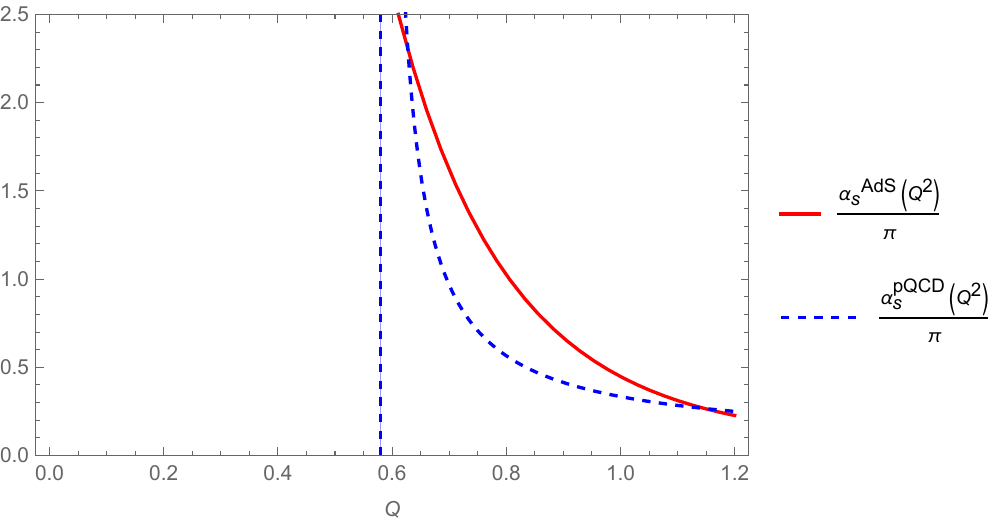}
\caption{Comparison of the running coupling from the UV deformation (red solid line) with the leading-order pQCD result (blue dashed line) for pure gluon states ($n_f = 0$). The Landau poles are matched at $\Lambda_{\mathrm{QCD}} = 0.58\,\mathrm{GeV}$ (vertical lines). The model shows a power-law falloff, whereas pQCD exhibits logarithmic running; this difference is expected, as the UV deformation describes the intermediate energy regime where the transition from nonperturbative to perturbative dynamics occurs. Reproducing the asymptotic logarithmic behavior would require bulk quantum corrections beyond this leading-order holographic framework.}
   \label{Figd}
\end{figure}

\subsection{Beta-Function} \label{DA2} 
The $\beta$-function is calculated using the renormalization group theory, 
\begin{align}\label{16}
\beta\!\left(\alpha^{\text{AdS}}(Q^2)\right) 
&= \frac{d\,\alpha^{\text{AdS}}(Q^2)}{d\ln Q^2} \nonumber\\
&= -\,\frac{Q^2 \pi \sigma}{2|\tilde{m}_\phi^8|}\left(f_\alpha^2 |\tilde{m}_\phi^2|\right)^{\tfrac{Q^2}{2|\tilde{m}_\phi^2|}}
\csc\!\left(\frac{\pi Q^2}{2|\tilde{m}_\phi^2|}\right)
\Bigg[
(2|\tilde{m}_\phi^2| - Q^2)
\Bigg(
\pi \cot\!\left(\frac{\pi Q^2}{2|\tilde{m}_\phi^2|}\right)
\nonumber\\
&\hspace{6cm} - \log\!\left(f_\alpha^2 |\tilde{m}_\phi^2|\Bigg)
\right) 
+\, 2|\tilde{m}_\phi^2|
\Bigg].
\end{align}
{The $\beta$-function follows from differentiating \cref{15}. It inherits the poles of $\alpha_s^{\mathrm{AdS}}(Q^2)$ (see \cref{Fig1}) at $Q^2 = 2n\tilde{m}_\phi^2$, where  $n$ is an integer, but remains well-defined and negative below the first pole ($Q^2 < 2 |\tilde{m}_\phi^2|$), satisfying asymptotic freedom.} In QCD, the $\beta$-function vanishes in both the UV (asymptotic freedom) and IR \cite{beta-f2} (conformal fixed-point \cite{beta-f1}) limits, reflecting the decoupling of colored degrees of freedom in the infrared \cite{Brodsky2}. 
The other condition is,
\begin{equation}\label{18}
\beta(Q)<0 \qquad\text{for}\qquad Q>0.
\end{equation} 
This condition reflects the anti-screening behavior of the QCD theory at large $Q$ where its running coupling, $\alpha_s$, vanishes due to asymptotic freedom with quark-gluon degrees of freedom. The QCD $\beta$-function is generally negative, and its value increases from the UV regime and assumes its maximum value in the IR domain.

 \begin{figure}[H]
  \centering
  \subfloat[Left Panel]{\includegraphics[scale=0.5]{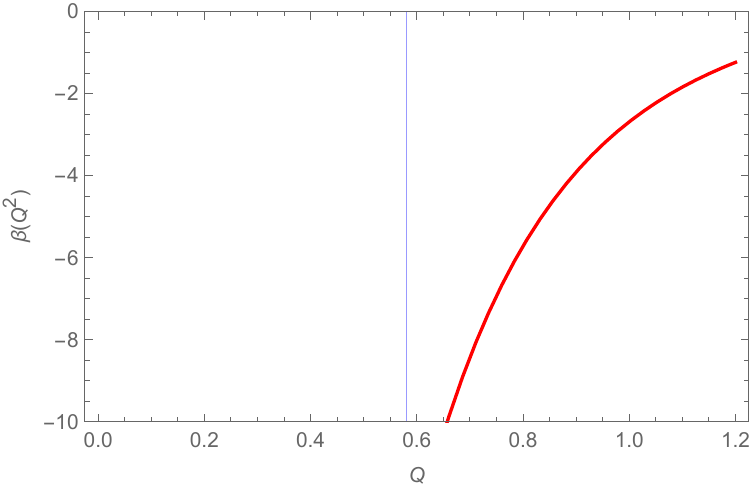}}
  \qquad
 \subfloat[Right Panel]{\includegraphics[scale=0.5]{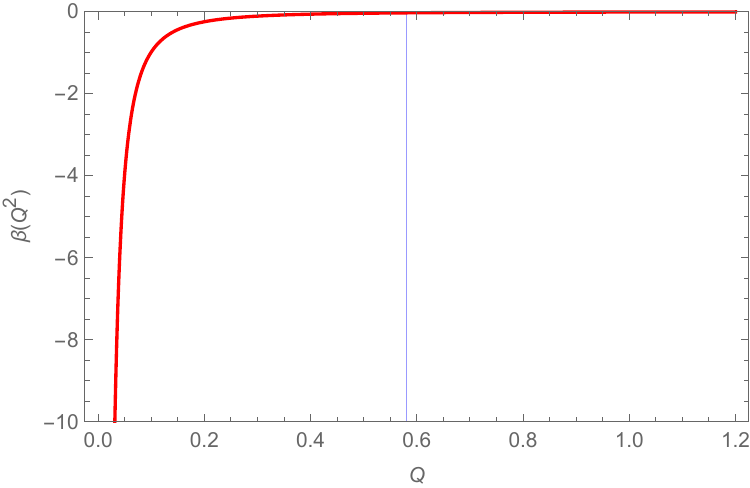}}
 \caption{A graph of $\beta(Q^2)$ against $Q$ for $|\tilde{m}_\phi|\,=\,0.86\,\text{GeV}$ (left) and $|\tilde{m}_\phi|\rightarrow\infty$ (right). The graph (left) is compared with $\beta(Q)$ (pQCD) reproduced in Refs.~\cite{Brodskya,ads-beta-f}. The graph in the right panel shows the behavior of the $\beta(Q^2)$ for $|\tilde{m}_\phi|\,\rightarrow\,\infty$.}
   \label{Figb}
\end{figure}
We expect the $\beta$-function to have a usual {analytical behavior} in both the UV and the IR regimes with or without the Landau pole in $\alpha_s$. The 'usual' behavior of $\beta(Q^2)$ is due to its conformal invariant characteristics, leading to the well-behaved cut-off in both the IR and the UV regimes. Regardless of the singularity exhibited by $\alpha_s$.

{The result in Eq.(\ref{15}) {displays the expected decrease in $\alpha_s$ at large $Q^2$} similar to the running coupling displayed for pQCD in \cref{Figd}. We determine the position of the Landau pole $Q_\Lambda$ in Eq.(\ref{15}) using the relation for determining the transition point, 
\begin{equation}\label{ceq}
    \dfrac{d\beta}{dQ}=0.
\end{equation}
However, the results obtained from the above operation are not analytically solvable for $Q_\Lambda$ because of the trigonometric functions involved. So we expand it for a small $Q$, leading to 
\begin{equation}\label{plnm}
    \dfrac{8}{|\tilde{m}^2_\phi| Q^3}+\dfrac{\left(\pi^2-3\log[f_\alpha^2|\tilde{m}^2_\phi|]\right)Q}{3|\tilde{m}_\phi^6|}+{\cal O}(Q)^3=0.
\end{equation}
This result yields,
\begin{equation}\label{tmm}
    Q_\Lambda=\left(\dfrac{24}{-\pi^2+\log[f_\alpha^2\tilde{m}^2_\phi]}\right)^{1/4}\tilde{m}_\phi.
\end{equation}
The apparent singularity from $\csc(\pi Q^2/(2 |\tilde{m}_\phi^2|))$ at $Q^2 = 2 |\tilde{m}_\phi^2|$ is canceled by the factor $(2 |\tilde{m}_\phi^2| - Q^2)$, so the actual Landau pole must be determined from the full expression, e.g., via $d\beta/dQ = 0$. Expanding this equation in the small parameter $Q/|\tilde{m}_\phi|$ (the relevant dimensionless ratio) gives the polynomial \cref{plnm}. Although $Q_\Lambda/|\tilde{m}_\phi| \approx 0.67$ is not negligible, $(Q_\Lambda/|\tilde{m}_\phi|)^2 \approx 0.45 < 1$, suppressing higher-order terms and making the first nonvanishing approximation reliable. The resulting $Q_\Lambda$ also agrees with the QCD scale from pQCD matching in \cref{Figd}, supporting the estimate. Using the parameters from Figs.~\ref{Fig1}, \ref{Figd}, and \ref{Figb}, $|\tilde{m}_\phi| = 0.86~\mathrm{GeV}$ and $f_\alpha = 0.50~\mathrm{GeV}^{-1}$, we find $Q_\Lambda \approx 0.58~\mathrm{GeV}$ for the Landau pole, indicated by a solid blue line in those figures. This singularity marks the scale where pQCD breaks down and is often associated with the confinement or hadronic scale. It also reflects the scale dependence of $\alpha_s$ and serves as an expansion parameter for related quantities such as the $\beta$-function~\cite{coupling}. The tachyon decay constant $f_\alpha = 1/\alpha = 0.50~\mathrm{GeV}^{-1}$ sets the scale of the tachyon field in the holographic direction. It is chosen so that the model's Landau pole $Q_\Lambda$ aligns with the QCD scale $\Lambda_{\rm QCD} \approx 0.58~\mathrm{GeV}$, ensuring a smooth UV--IR transition (\cref{Figd}).  To verify the reliability of the expansion, we numerically solved the full condition ${d\beta}/{dQ} = 0$ using the exact expression for \(\beta(Q)\) (\cref{16}). With the parameters \(|\tilde{m}_\phi| = 0.86~\mathrm{GeV}\) and \(f_\alpha = 0.50~\mathrm{GeV}^{-1}\), the root is found at $Q_\Lambda = 0.58273~\mathrm{GeV}$, in excellent agreement with the approximate value \(0.58~\mathrm{GeV}\). This confirms the validity of the truncated expansion.

\section{Deformation of the AdS space in the IR regime}\label{D}
For the nonperturbative behavior of couplings, we perturb the tachyon fields about their 'true' vacuum $\phi_0\,=\,\pm a$ by introducing a new field $\eta$ through $\phi\,\rightarrow\,\eta\,+\,\phi_0$. This stabilizes the tachyons and the new field acquires a positive mass $V''(\phi)\vert_{\phi_0=\pm a}\,=\,2a^2\Lambda\,=\,m_\phi^2$. Consequently, Eq.(\ref{13}) becomes
\begin{align}\label{n}
    V(\eta)&=V(\phi)\big\vert_{\phi_0}+V'(\phi)\big\vert_{\phi_0}\eta+\dfrac{1}{2}V''(\phi)\big\vert_{\phi_0}\eta^2\nonumber\\
    &=\dfrac{1}{2}m_\phi^2\eta^2,
\end{align}
this process coincides with \textit{tachyon condensation}. The new field $\eta$ is identified with a stable glueball field \cite{Issifu:2022pif} in the IR regime with a solution
{\begin{equation}\label{n4}
    \eta(\zeta)=\dfrac{ e^{m_\phi \zeta}}{\zeta},
\end{equation}}
 which produces confinement of glueballs in the limit of large particle separation distance, where \textit{color confinement} is expected. The solution, \(\eta(\zeta)\), is obtained by including the kinetic term \(\frac{1}{2}(\partial_\mu \phi)^2\) of the tachyonic field in the action (presented in \cref{tb}) and expanding around the true vacuum \(\phi_0 = \pm a\) with the shift \(\phi = \phi_0 + \eta\) \cite{Issifu3}. The linearized equation of motion in spherical coordinates is given by:
 \[
\Big[-\frac{d^2}{d\zeta^2} + \frac{2}{\zeta^2} + V''(\phi_0)\Big] \eta(\zeta) = 0,
\]
 with $m_\phi^2 = V''(\pm a) = 2 \Lambda a^2$, which corresponds to the physical scalar glueball mass. This procedure follows the standard holographic treatment of scalar fields in AdS \cite{correspondence-a, correspondence-b, Klebanov, Megias:2010ku} and is consistent with the AdS/CFT dictionary \cite{correspondence-a, correspondence}. A comprehensive study can be found in our previous papers \cite{Issifu, Issifu1, Issifu2, Issifu3}, further establishing that in the limit, $r\,\rightarrow\,\infty$, this solution is dominant.
 
 \subsection{Strong Running Coupling}\label{D1}
 Using the \textit{color dielectric function} for the condensed tachyons, Eq.(\ref{12a}) alters to
\begin{equation}
    \alpha_s^{AdS}(\zeta)\propto G(\eta(\zeta))^{-1}.
\end{equation}
Using the Laplace transform
\begin{align}\label{n1}
    \alpha_s^{AdS}(Q^2)&\sim \int_0^\infty\alpha_s^{AdS}(\zeta)e^{-Q^2\zeta/m_\phi}d\zeta\nonumber\\
    &=\dfrac{4m_\phi\rho}{(2m_\phi^2+Q^2)^3},
\end{align}
where $\rho$ is a proportionality constant. At $Q\,=\,0$, $\rho\,=\,2m_\phi^5\alpha_s^{AdS}(0)$, and the above expression transforms into
{\begin{equation}\label{n2}
\alpha_s^{AdS}(Q^2) =\dfrac{(2m_\phi^2)^3}{(2m_\phi^2+Q^2)^3}\alpha^{AdS}_s(0).
\end{equation}}
 The condensed phase \(\eta(\zeta)\) yields a confining background. The color dielectric function \(G(\eta)\) produces a linear confining potential (area law) \cite{dePaula:2008fp} and a mass gap \(m_\phi\) \cite{Issifu:2022pif, Afonin:2010fr}. By contrast, the UV phase with free tachyons corresponds to a deconfined, perturbative regime (discussed in \cref{DA}). Thus, the model provides a clear distinction between confining and non-confining scenarios through the vacuum expectation value of the tachyon field.

   \begin{figure}[H]
  \centering
  \subfloat[Left Panel]{\includegraphics[scale=0.5]{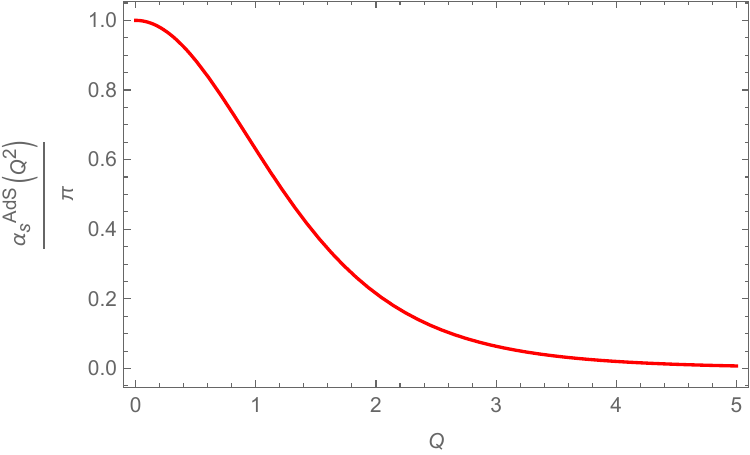}}
  \qquad
  \subfloat[Right Panel]{\includegraphics[scale=0.5]{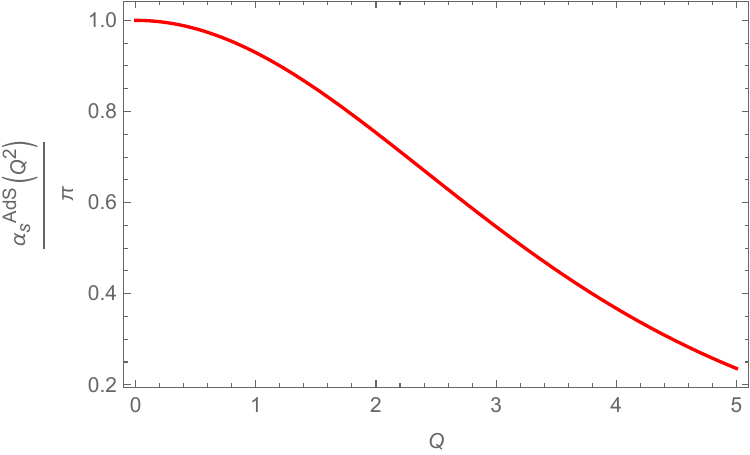}}
  \caption{Left panel: $\alpha_s^{\mathrm{AdS}}(Q^2)/\pi$ as a function of $Q$ for the physical glueball mass $m_\phi = 1.73\,\mathrm{GeV}$ (from lattice QCD). The solid red curve is compared with the effective charge $\alpha_{s,g1}(Q)$ extracted from the Bjorken sum rule (data points from Refs.~\cite{Brodskya, AdS-book, ads-beta-f}), normalized such that $\alpha_s^{\mathrm{AdS}}(0) = \pi$. All model parameters are fixed by independent lattice and QCD inputs (see Secs.~\ref{DA}--\ref{D}), so the agreement constitutes a genuine test of the framework. Right panel: the same coupling in the limiting case $m_\phi \to \infty$ (artificially heavy glueballs), illustrating that increasing $m_\phi$ leads to stronger binding, as expected within the model.}
   \label{cn}
\end{figure}

\subsection{Beta-Function}\label{D2}
 We can also calculate the $\beta$-function from Eq.(\ref{16}) which leads to the expression 
 {\begin{equation}\label{n3}
     \beta(Q^2)=-\dfrac{3(2m_\phi^2)^3 Q^2}{(2m_\phi^2+Q^2)^4}\alpha^{AdS}_s(0).
 \end{equation}}
 This expression satisfies the restrictions 
 \begin{equation}
     \beta(Q\rightarrow 0)=\beta(Q\rightarrow \infty)=0\qquad\text{and}\qquad \beta(Q)<0 \qquad\text{for}\qquad Q>0.
 \end{equation}
 The first condition satisfies the restrictions that QCD approximates conformal theory in both the far UV and the deep IR regimes, known as "the principle of maximum conformality":  meanwhile, the second condition is a consequence of the asymptotic freedom properties of QCD theory. {Hence, $\alpha_s^{AdS}(Q^2)$ can be normalized to $\alpha_{s,g_1}$ (see \cref{cn} above), a well-measured effective charge obtained from the Bjorken sum rule \cite{ Bjorken} for polarized deep inelastic lepton-proton scattering, by exploring the similarities in the qualitative behavior of the $\beta$-functions. It vanishes in the deep infrared domain, rises to a higher negative immediately, and fall-off to zero in the far UV domain. We take advantage of these similar features and normalize $\alpha_s^{AdS}(Q^2=0)=\pi$ for comparison.}
 Taking the derivative with respect to $Q$
 {\begin{equation}
     \dfrac{d\beta}{dQ}=-\dfrac{6(2m_\phi^2)^3Q\pi}{(2m^2_\phi+Q^2)^4}\left[1-\dfrac{4Q^2}{2m^2_\phi+Q^2}\right],
 \end{equation}}
 consequently, there is a transition at $Q_0\,=\,\sqrt{(2/3)}m_\phi$. Perturbation theory is applicable in the regime $Q_0\,>\,\sqrt{(2/3)}m_\phi$ whilst non-perturbation theory is applicable in the regime $Q_0\,<\,\sqrt{(2/3)}m_\phi$. Also, the conditions 
 \begin{equation}
     \dfrac{d\beta}{dQ}\Big\vert_{Q=Q_0}=0,\qquad\qquad  \dfrac{d\beta}{dQ}\Big\vert_{Q_0}<0\quad \text{for}\quad Q<Q_0\qquad\text{and}\qquad \dfrac{d\beta}{dQ}\Big\vert_{Q_0}>0 \quad \text{for}\quad Q>Q_0
 \end{equation}
are all satisfied \cite{Brodsky1, Brodskya}.

 \begin{figure}[H]
  \centering
  \subfloat[Left Panel]{\includegraphics[scale=0.4]{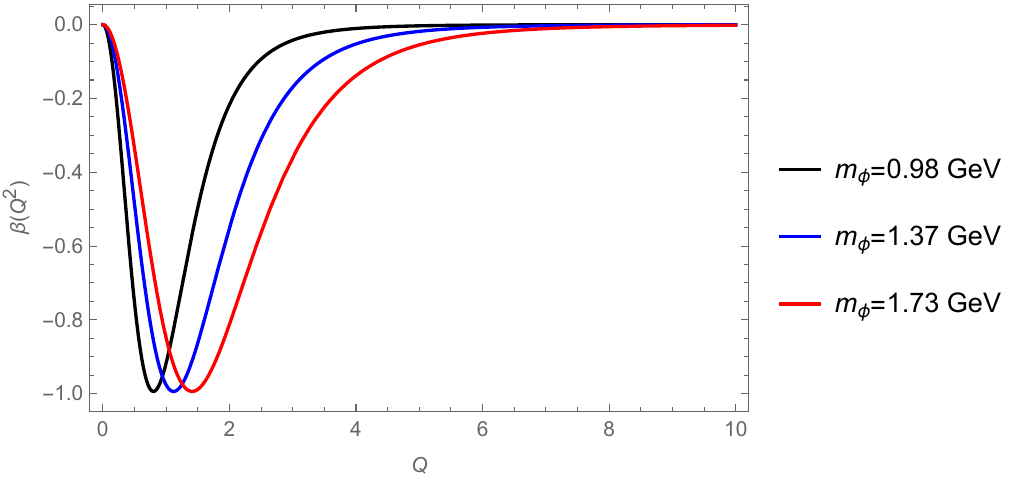}}
  \qquad
  \subfloat[Right Panel]{\includegraphics[scale=0.4]{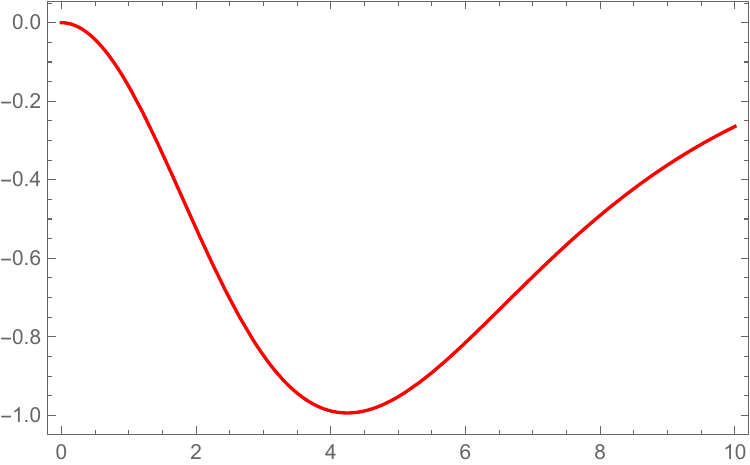}}
  \caption{Left panel: $\beta(Q^2)$ as a function of $Q$ for three glueball masses $m_\phi$ (colored curves). All parameters are fixed by independent lattice inputs (see Secs.~\ref{DA}--\ref{D}). The transition momentum $Q_0 = 2/3\,m_\phi$ marks the minimum of $\beta(Q)$: for $m_\phi = 1.73$~GeV (red), $Q_0 \approx 1.41$~GeV; for $m_\phi = 1.37$~GeV (black), $Q_0 \approx 1.12$~GeV; for $m_\phi = 0.98$~GeV (blue), $Q_0 \approx 0.80$~GeV. The curves are compared with the AdS/QCD and modified AdS/QCD $\beta$-functions from Refs.~\cite{Brodskya,ads-beta-f}, showing better agreement with the modified AdS case. For $Q > Q_0$, $\beta(Q)$ remains negative, reflecting asymptotic freedom. Right panel: the limiting case $m_\phi \to \infty$ (all curves collapse to the same shape) illustrates that heavier glueballs shift the transition to larger momenta, consistent with the scaling $Q_0 \propto m_\phi$. 
  }
   \label{Fig2c}
\end{figure}

\begin{figure}[H]
  \centering
 {\includegraphics[scale=0.7]{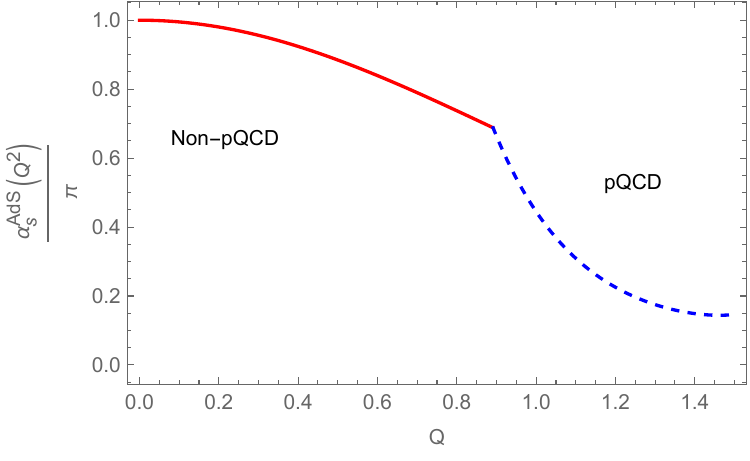}}
\caption{A unified diagram for $\alpha_s^{AdS}(Q^2)$ in both the IR (non-pQCD) and the UV (pQCD) regions. A schematic diagram showing the behavior of the strong running coupling in both the UV (dashed blue) and IR (red) regions in the model framework. This diagram shows the transition point at $0.89\,\text{GeV}$ (the two graphs \ref{Fig1} (left) and \ref{cn}(left) intersect at this point), corresponding to $m_\phi\,\approx\,1.09\,\text{GeV}$.}
   \label{cp}
\end{figure}

\section{Analysis and Conclusion}\label{A}
\subsection{Analysis}\label{A1}
The graphs plotted for the couplings determined from the UV region were for $f_\alpha\,=\,0.50\,\text{GeV}$ and $|\tilde{m}_\phi|\,=\,0.86\,\text{GeV}$, both determined via the Landau pole condition. The glueball mass $m_\phi$ is fixed by the tachyon potential: expanding around $\phi_0 = \pm a$ gives $m_\phi^2$. We calibrate it to match the lattice QCD value $m_\phi = 1.73\ \mathrm{GeV}$, preserving the model's predictive power. Generally, we considered the \textit{lightest scalar glueball mass} consistent with the background of the study presented in Sec.~\ref{SG}. From QCD lattice simulations results, QCD sum rule, and QCD phenomenology, the \textit{lightest scalar glueball mass} with quantum number $J^{PC}\,=\,0^{++}$ and frequency $f_0(1710)$, has been determined to be $m_\phi\,=\,1.73\,\text{GeV}$ \cite{Issifu1, Issifu2, Issifu3}(and references therein). Other glueball masses with the same quantum number and frequencies $f_0(500),\,f_0(980),\,f_0(1370)$ have also been identified \cite{Ochs, Crede}. {\cref{Fig2c} shows how the transition point between the two regimes, marked by $\beta(Q^2)$, depends on the magnitude of $m_\phi$.}

We emphasize that the expression in \cref{15} is valid for $Q^2 < 2|\tilde{m}_\phi^2|$, where it remains real and well defined. In our analysis, the Landau pole at $Q_\Lambda \approx 0.58\,\text{GeV}$ lies well below the threshold $Q \approx \sqrt{2}\,|\tilde{m}_\phi|\approx 1.22\,\text{GeV}$, so the model applies to the intermediate energy region rather than the asymptotic UV. For $Q^2 \gg 2|\tilde{m}_\phi^2|$, the expression develops poles and ceases to be physical. At large $Q^2$, Eq.~(22) exhibits a power-law (exponential) falloff, not the logarithmic running of pQCD, confirming that the UV deformation captures the transition from the IR to intermediate scales. Reproducing asymptotic behavior would require bulk quantum corrections beyond this leading-order framework.

 We established the connection between the couplings ($\alpha_s^{\rm AdS}(Q^2)$ and $\beta(Q^2)$) and the mass scales $m_\phi$ and $\tilde{m}_\phi$ in the IR and UV regions respectively. The Landau pole is located at $Q_\Lambda \approx 0.58~\text{GeV}$ (\cref{tmm}), related to the UV mass scale, while the transition momentum in the IR is $Q_0 = \sqrt{2/3}\,m_\phi$, giving $Q_0 \approx 1.41~\text{GeV}$ for $m_\phi = 1.73~\text{GeV}$. This indicates that heavier glueballs are more tightly bound, analogous to the behavior of heavy quarks. The UV mass scale \(|\tilde{m}_{\phi}| = 0.86~\mathrm{GeV}\) and the IR glueball mass \(m_\phi = 1.73~\mathrm{GeV}\) both arise from the tachyon potential \cref{13}. Expansions about the false and true vacua give \(\tilde{m}_\phi^2=-\Lambda a^2\) and \(m_\phi^2=2\Lambda a^2\) respectively, implying \(m_\phi=\sqrt{2}\,|\tilde{m}_\phi|\). Analogous to connections between the scalar glueball mass and the gluon condensate found in QCD effective models \cite{Cornwall}. While the lattice value \(m_\phi=1.73~\mathrm{GeV}\) predicts \(|\tilde{m}_\phi|\approx1.22~\mathrm{GeV}\), we fix \(|\tilde{m}_{\phi}|=0.86~\mathrm{GeV}\) phenomenologically by matching the Landau pole to \(\Lambda_{\mathrm{QCD}}\), with the \(\sim30\%\) deviation consistent with leading-order accuracy. The transition scale \(Q_0\approx0.89~\mathrm{GeV}\) (\cref{cp}) then emerges as the intersection of the UV and IR couplings.
 
 The power‑law falloff in \cref{n2} characterizes the confining phase and describes the coupling in the infrared and intermediate momentum regions (up to a few GeV), remaining finite at $Q=0$. It does not reproduce the asymptotic logarithmic running of pQCD, which occurs in the deep UV and is governed by the free‑tachyon deformation with its unphysical pole at $Q=0$ (\cref{DA1}). Together, the IR and UV sectors provide a unified description across scales, with the transition near \(Q_0 \sim 0.89~\text{GeV}\) (\cref{cp}).

This paper presents a new approach to determining the AdS/QCD running coupling and associated $\beta$-function. Rather than fixing $\alpha_s^{AdS}(Q^2)$ in the poorly understood IR and extrapolating to the UV, we introduce a tachyon-induced function $G(\phi)$ that deforms AdS space differently in the free (UV) and condensed (IR) regimes. The resulting couplings are compared with the effective charge $\alpha_{s,g1}$ from the Bjorken sum rule.

In the UV, we obtain a sinusoidal behavior, $\alpha_s^{AdS}(Q^2)\sim 1/\sin(\pi Q^2/2|\tilde{m}_\phi^{2}|)$, in contrast to the logarithmic pQCD running $\alpha_s(Q^2)\sim 1/\ln(Q^2/\Lambda^2_{\text{QCD}})$. In the IR, instead of the exponential soft-wall result $\alpha_s^{AdS}(Q^2)\sim e^{-Q^2/4\kappa^2}$, our model yields a power-law falloff $\alpha_s^{AdS}(Q^2)\sim 1/(2|\tilde{m}_\phi^{2}|+Q^2)^3$ (see \cref{cn}). Given that the IR behavior of the QCD coupling is not uniquely determined experimentally~\cite{gp1,Deur}, hadronic observables remain the most reliable benchmarks for comparison with lattice results~\cite{coupling1}. Finally, the derived $\beta$-functions satisfy the principle of maximum conformality in both UV and IR regimes.

The UV deformation of AdS space (Sec.~\ref{DA}) leads to a nonphysical \textit{Landau singularity} in $\alpha_s^{AdS}(Q^2)$ at $Q_\Lambda$, signaling the breakdown of perturbative QCD. In contrast, the IR deformation (Sec.~\ref{D}) yields a regular coupling in both the deep IR and far UV, consistent with the expected \textit{IR freezing}. Experimental~\cite{coupling-x} and lattice~\cite{coupling-l} results support a finite coupling across all scales, motivating mechanisms that remove the Landau pole and IR renormalons~\cite{confinement-1} and ensure finite $Q$.

At low energies ($Q^2\to 0$), gluons are expected to acquire a dynamical mass, $Q^2=-q^2\cong m_A^2$~\cite{g-mass,Badalian:2003cm,Badalian:1999fq,IR-fix}, which regulates the IR behavior and leads to a finite coupling. Introducing $m_A^2$ into $Q^2$ preserves the UV limit while dominating in the IR, thereby defining an \textit{IR freezing point}~\cite{IR-freezing-ph,Badalian:2004ig,IR-freezing-lt,Badalian:2000hv,beta-f2} and improving the description of confinement. This dynamically generated gluon mass is unrelated to chiral symmetry breaking; instead, it reflects strong gluon interactions that form bound states such as scalar ($J^{PC}=0^{++}$) glueballs, associated with the gluon condensate $\langle G^{\mu\nu}G_{\mu\nu}\rangle$.

In this context, the gluon mass can be related to the scalar glueball mass. For instance, leading-order Yang--Mills analyses with an auxiliary field $\phi$ give $m_A = m(0^{++})/\sqrt{6}$~\cite{Cornwall,Kondo}, while color dielectric models yield $m_A = m_\phi/2$~\cite{Issifu2,Issifu3}. Accordingly, $\alpha_s^{AdS}(Q^2)$ in the UV region takes the form
\begin{equation}\label{15a}
\alpha^{AdS}_s(Q^2+m_A^2)=\dfrac{\sigma\pi}{|\tilde{m}_\phi^{6}|}\left({|\tilde{m}_\phi^{2}|f_\alpha^2}\right)^{(Q^2+m_A^2)/2|\tilde{m}_\phi^{2}|} (2|\tilde{m}_\phi^{2}|-(Q^2+m_A^2))\csc\left(\dfrac{\pi (Q^2+m_A^2)}{2|\tilde{m}_\phi^{2}|}\right).
\end{equation}
For $|\tilde{m}_\phi|\,=\,0.86$ we get $m_A\,=\,0.43$ which leads to a finite IR freezing point as presented in \cref{Fig2} and \cref{Fig2a}. On the other hand, the IR model self-regulates at low $Q^2$ with a finite freezing point normalized at $\alpha^{AdS}_s(0)\,=\pi$, on $g_1$ scheme \cite{AdS-book}. 
\begin{figure}[H]
  \centerline{\includegraphics[scale=0.8]{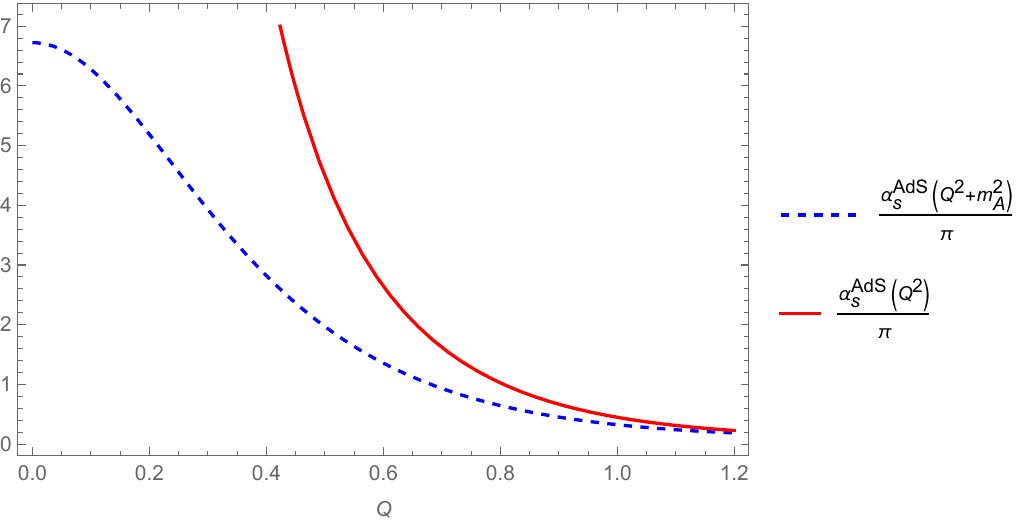}}
  \caption{A graph of $\alpha_s^{AdS}(Q^2+m_A^2)/\pi$ (dashed blue) and $\alpha_s^{AdS}(Q^2)/\pi$ (red) against $Q$ for $m_A=0.43\,\text{GeV}$ and $|\tilde{m}_\phi|=0.86\,\text{GeV}$. The dashed blue graph is well-behaved in the IR region due to the IR freezing point, while the red graph shows the singularity in the UV region. However, in the limit of high $Q$, they both fall off in a similar manner.}
  \label{Fig2}
\end{figure}

\begin{figure}[H]
  \centerline{\includegraphics[scale=0.8]{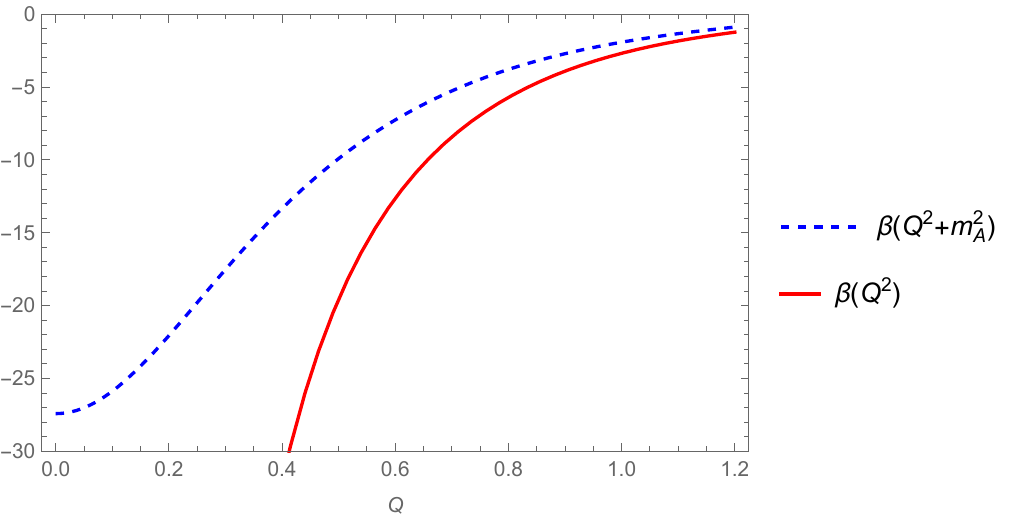}}
  \caption{A graph of $\beta(Q^2+m_A^2)$ (dashed blue) and $\beta(Q^2)$ (red) against $Q$ for $m_A=0.43\,\text{GeV}$ and $|\tilde{m}_\phi|=0.86\,\text{GeV}$. We compare $\beta$-function with IR freezing to the one without IR freezing.}
  \label{Fig2a}
\end{figure}

\subsection{Conclusion}\label{sec3}
In this work, we constructed the tachyonic AdS/QCD action from the DBI action and studied the running coupling $\alpha_s^{AdS}(Q^{2})$ and its associated $\beta(Q^{2})$ by distorting the $\text{AdS}_5$ background through tachyon-induced color dielectric functions. We showed that free tachyons deform the AdS space in the UV region, reproducing qualitative features consistent with perturbative QCD and the $\alpha{s,g1}$ effective charge, while condensed tachyons dominate in the IR, yielding behavior characteristic of nonperturbative QCD. Increasing the glueball mass $m_\phi$ enhances color confinement and raises the transition momentum, with heavier glueballs confining more strongly than lighter ones.

We further provided a unified description of $\alpha_s^{AdS}(Q^{2})$ across UV and IR scales, identifying a transition at $Q_0 \sim 0.89,\text{GeV}$ corresponding to $m_\phi \approx 1.09,\text{GeV}$. A Landau singularity was observed in the UV, which we resolved by introducing a dynamically generated gluon mass, thereby correcting the unphysical $Q \to 0$ limit. Overall, our framework establishes a consistent relation between $\alpha_s^{AdS}(Q^{2})$, $\beta(Q^2)$, and $m_\phi$ ($|\tilde{m}_\phi|$), showing that suitable tachyon potentials, particularly of Higgs-like form, allow the AdS background to capture both perturbative and nonperturbative QCD dynamics within a unified approach.

{The framework developed in this work opens several promising directions for future investigation. The use of a tachyon-induced color dielectric function to deform the gravitational background is not restricted to the AdS/QCD context considered here. It would be of particular interest to extend this mechanism to other gauge/gravity dualities, including non-supersymmetric AdS/CFT scenarios, to study confinement–deconfinement transitions in a broader setting. Additionally, exploring a possible AdS/Electroweak correspondence may offer new insights into the running of the electroweak coupling and mass generation, in analogy with the holographic role played by the tachyon in this model. Finally, a direct phenomenological application of the present framework lies in fitting the glueball-mediated potential derived here to heavy quarkonium spectroscopy, which would allow a quantitative assessment of glueball contributions to quark confinement across different mass scales.}

{
\appendix
\section{Detailed Derivation of the Mellin Transform}
\label{app:mellin}
The running coupling is defined through the Mellin transform
\begin{equation}
\alpha_s^{\mathrm{AdS}}(Q^2) \propto \int_0^\infty 
\alpha_s^{\mathrm{AdS}}(\zeta)\,\zeta^{s-1}\,d\zeta,
\qquad 
s \equiv \frac{Q^2}{|\tilde{m}_\phi^2|},
\end{equation}
with $\alpha_s^{\mathrm{AdS}}(\zeta) \propto [(\alpha\zeta)^2 - a^2]^{-2}$.
The integrand has a double pole at $\zeta_0 = a/\alpha$ on the real axis,
regulated by $a^2 \to a^2 - i\varepsilon$ ($\varepsilon \to 0^+$).

\subsection*{Dimensionless form}

Introducing $u = \zeta/\zeta_0$, so $\zeta = \zeta_0 u$, $d\zeta = \zeta_0\,du$. Then
\begin{equation}
(\zeta\alpha)^2 - a^2 = \alpha^2\zeta_0^2(u^2-1) = a^2(u^2-1),
\end{equation}
and the integrand becomes
\begin{equation}
\alpha_s^{\mathrm{AdS}}(Q^2) \propto \frac{\zeta_0^s}{a^4}\,I(s),
\qquad
I(s) = \int_0^\infty \frac{u^{s-1}}{(u^2-1-i0^+)^2}\,du.
\end{equation}

\subsection*{Integration by parts}
Write $(u^2-1)^{-2} = (u-1)^{-2}(u+1)^{-2}$ and integrate by parts with $dv = -(d/du)(u-1-i0^+)^{-1}\,du$ and $w = u^{s-1}(u+1)^{-2}$:
\begin{equation}
I(s) = \left[\frac{-u^{s-1}}{(u+1)^2(u-1-i0^+)}\right]_0^\infty
+ \int_0^\infty \frac{f'(u)}{u-1-i0^+}\,du,
\qquad f(u) = \frac{u^{s-1}}{(u+1)^2}.
\end{equation}
The boundary term vanishes for $0 < \mathrm{Re}(s) < 2$: at $u\to\infty$ it decays as $u^{s-3}\to 0$; at $u=0$ it vanishes as $u^{s-1}\to 0$ for $s>0$. This range covers the entire physical domain $0 < Q^2 < 2|\tilde{m}^2_\phi|$.

\subsection*{Sokhotski--Plemelj decomposition}
With $f'(u) = (s-1)u^{s-2}(u+1)^{-2} - 2u^{s-1}(u+1)^{-3}$, the Sokhotski--Plemelj theorem gives
\begin{equation}
I(s) = \mathrm{P.V.}\int_0^\infty \frac{f'(u)}{u-1}\,du + i\pi f'(1), \qquad f'(1) = \frac{s-2}{4}.
\end{equation}
The imaginary part $i\pi(s-2)/4$ is real-valued and nonzero for $s\neq 2$. It is discarded by the physical reality condition: $\alpha_s^{\mathrm{AdS}}(Q^2)$ is defined as the real part of the analytically continued expression, and $i\pi(s-2)/4$ is a contact term that vanishes at the boundary of the physical domain $s\to 2$ (i.e.\ $Q^2\to 2|\tilde{m}^2_\phi|$) where the coupling reaches its first pole.

\subsection*{Partial fraction decomposition}
To apply the Mellin identity to the principal value integral, decompose via partial fractions:
\begin{equation}
\frac{f'(u)}{u-1} = \frac{(s-1)u^{s-2}}{(u-1)(u+1)^2} - \frac{2u^{s-1}}{(u-1)(u+1)^3}.
\end{equation}
Using
\begin{align}
\frac{1}{(u-1)(u+1)^2} &= \frac{1}{4(u-1)} - \frac{1}{4(u+1)} - \frac{1}{2(u+1)^2}, \\
\frac{1}{(u-1)(u+1)^3} &= \frac{1}{8(u-1)} - \frac{1}{8(u+1)} - \frac{1}{4(u+1)^2} - \frac{1}{2(u+1)^3},
\end{align}
and the substitution $u\to 1/u$ to relate integrals over $(u+1)^{-k}$ to standard Mellin integrals, all terms reduce to combinations of 
$\int_0^\infty u^{q-1}/(u+1)^k\,du$, which are evaluated by 
$B(q,k-q) = \Gamma(q)\Gamma(k-q)/\Gamma(k)$.

\subsection*{Final result}

Collecting all terms and applying the Euler reflection formula 
$\Gamma(p)\Gamma(1-p) = \pi/\sin(\pi p)$ yields
\begin{equation}
\mathrm{P.V.}\int_0^\infty \frac{f'(u)}{u-1}\,du 
= \frac{\pi(2-s)}{4}\,\csc\!\left(\frac{\pi s}{2}\right),
\end{equation}
so that
\begin{equation}
\alpha_s^{\mathrm{AdS}}(Q^2)
\propto (2|\tilde{m}_\phi^2| - Q^2)\,
\csc\!\left(\frac{\pi Q^2}{2|\tilde{m}_\phi^2|}\right),
\end{equation}
where the factor $(2-s) = (2|\tilde{m}^2_\phi|-Q^2)/|\tilde{m}^2_\phi|$ comes directly from the residue $f'(1) = (s-2)/4$ (via the $(u-1)^{-1}$ terms in the partial fraction decomposition) and the $\csc$ from the Euler reflection structure of the Beta function. 

}

\section*{Glossary of Abbreviations and Terms}
\begin{itemize}
    \item \textbf{QCD} — Quantum Chromodynamics, the fundamental theory describing strong interactions among quarks and gluons.
    \item \textbf{pQCD} — Perturbative QCD, the high-energy expansion of QCD valid at short distances.
    \item \textbf{UV} — Ultraviolet; high-energy or short-distance regime.
    \item \textbf{IR} — Infrared; low-energy or long-distance regime.
    \item \textbf{\(\Lambda_{\text{QCD}}\)} — QCD scale parameter separating perturbative and nonperturbative regimes.
    \item \textbf{YM Theory} — Yang–Mills gauge theory underlying QCD.
    \item \textbf{LF} — Light-Front formulation of relativistic dynamics on surfaces defined by \(x^+ = x^0 + x^3\).
    \item \textbf{LFWF} — Light-front wavefunction representing the quantum state of a hadron.
\end{itemize}

\subsection*{Holography and String Theory}
\begin{itemize}
    \item \textbf{AdS} — Anti-de Sitter spacetime with constant negative curvature.
    \item \textbf{CFT} — Conformal Field Theory; a quantum field theory invariant under conformal transformations.
    \item \textbf{AdS/CFT} — Duality between string/gravity theory in AdS space and a CFT on its boundary.
    \item \textbf{AdS/QCD} — Application of AdS/CFT techniques to model QCD phenomena.
    \item \textbf{SYM Theory} — Supersymmetric Yang–Mills theory, e.g., $\mathcal{N}=4$ SYM in holography.
    \item \textbf{DBI Action} — Dirac–Born–Infeld action describing D-brane dynamics.
    \item \textbf{D$p$-brane} — $p$-dimensional object in string theory where open strings can end.
    \item \textbf{IIB String Theory} — One of the two ten-dimensional supersymmetric superstring theories.
\end{itemize}

\subsection*{Model-Specific Quantities}
\begin{itemize}
\item \textbf{Color dielectric function} $G(\phi)$ — Dimensionless function induced by the tachyon field, used to deform the AdS geometry. Controls the behavior of the running coupling across UV and IR scales.
\item \textbf{Light-front holography} — Mapping between the AdS wave equation and the light-front Hamiltonian of QCD bound states.
\item \textbf{Landau pole / Landau singularity} — Scale at which a perturbative running coupling diverges, signaling breakdown of perturbation theory.
\end{itemize}

\acknowledgments
This work is partially supported by Conselho Nacional de Desenvolvimento
Científico e Tecnológico (CNPq) project No.: 168546/2021-3, Brazil. A. I. and E. A. A. acknowledge
financial support from the São Paulo State Research Foundation (FAPESP), Grants No. 2023/09545-1, 2025/17347-0,
and 2024/05377-0. F. A. B. acknowledges support from CNPq (Grant No. 309092/2022-1)
\\
\\
Data Availability Statement: The datasets used and/or analyzed during the current study are available from the corresponding author upon reasonable request.

 \bibliographystyle{unsrt}
\bibliography{references}


\end{document}